\documentclass[hyper]{JHEP3} 

\usepackage{graphicx}
\usepackage{epsfig}
\usepackage{cite}

\title{Collider Phenomenology with Split-UED}

\author{Kyoungchul Kong\\
        Theoretical Physics Department, SLAC, 
        Menlo Park, CA 94025, USA \\
        E-mail: \email{kckong@slac.stanford.edu}
        }

\author{Seong Chan Park\\
	Institute for the Physics and Mathematics of the Universe, \\
        The University of Tokyo, Kashiwa, Chiba 277-8568, JAPAN\\
	E-mail: \email{seongchan.park@ipmu.jp}
        }

\author{Thomas G. Rizzo\\
        Theoretical Physics Department, SLAC, 
        Menlo Park, CA 94025, USA \\
        E-mail: \email{rizzo@slac.stanford.edu}
        }

\preprint{
          IPMU10-0020 \\
          SLAC-PUB-13946 \\
          \today     }	

\abstract{
We investigate the collider implications of Split Universal Extra Dimensions.
The non-vanishing fermion mass in the bulk, 
which is consistent with the KK-parity, largely modifies 
the phenomenology of Minimal Universal Exta Dimensions.
We scrutinize the behavior of couplings and study the discovery reach of 
the Tevatron and the LHC for level-2 Kaluza-Klein modes 
in the dilepton channel, which would indicates the presence of the extra dimensions.
Observation of large event rates for dilepton resonances can result from 
a nontrivial fermion mass profile along the extra dimensions, which, in turn, may 
corroborate extra dimensional explanation for the observation of the positron excess in cosmic rays.
}

\keywords{Beyond Standard Model, Extra Dimension, Kaluza-Klein particle}

\begin{document}

\section{Introduction}
\label{sec:intro}

Models with extra spatial dimensions allow us to confront some of the 
outstanding issues of the Standard Model (SM) (see \cite{ADD, RS,ED6}).
In particular, the Universal Extra Dimensions (UED) scenario \cite{Appelquist:2000nn} 
leads to an interesting dark 
matter candidate \cite{Servant:2002aq,Cheng:2002ej,Kong:2005hn,Burnell:2005hm,Arrenberg:2008wy} 
as well as a foil for searches for Supersymmetry 
at colliders \cite{Rizzo:2001sd,Cheng:2002ab,Datta:2005zs,Battaglia:2005zf}. In the original minimal UED picture (MUED), all of the 
SM fields live in a 5-dimensional $S^1/Z_2$ orbifolded bulk with a 
compactification radius $R$. Due to the breaking of 5D Lorentz invariance, 
Kaluza-Klein (KK) number is no longer conserved although a $Z_2$ symmetry, 
KK-parity, remains. This being the case, the tree-level wave functions for the 
various KK states are either sines or cosines in the coordinate 
of the extra dimension.  Allowing for radiative loop 
corrections to the tree-level particle masses, the physics of MUED is then 
described by only two parameters beyond those of the SM \cite{Cheng:2002iz}: 
$R$ and a cutoff 
scale, $\Lambda$, used to define these loop corrections, which is usually 
taken such that $\Lambda R \sim {\cal O}(10-100)$ but with only logarithmic sensitivity 
to this particular choice.

In MUED and its extension to higher dimensions 
\cite{Burdman:2006gy,Dobrescu:2007xf,Freitas:2007rh,Dobrescu:2007ec}, 
the bulk masses of the 
SM fermions are taken to be zero. However, this is no longer the case in 
Split-UED (SUED) \cite{sued1,sued2,sued3,sued4}. Indeed this `bulk mass' term is naturally included in the effective Lagrangian as the term is compatible with 5D Lorentz invariance as well as gauge invariance of the model. Here one notes that in order to maintain the KK parity the `coefficient' of the $\bar \Psi \Psi$ fermion bilinear term in the action must be  an odd function of the 5D coordinate, 
$y$, defined on an interval, $y \in (-L, L)$ where $L=\frac{\pi R}{2}$. 
The simplest choice to make in 
this case, as is similarly done in the Randall-Sundrum (RS) model, 
is to write this coefficient as $\mu \theta(y)$, where $\mu$ is 
a dimensionful parameter whose value is, in general, dependent upon which 
SM field is being considered and $\theta(y)=1(-1)$ for $y>(<)0$. Naturally, 
one might expect that the values of $\mu$ can be of either sign and be of 
order $\sim 1/R$. The effects of including a non-zero value for $\mu$ are   
two-fold: First, depending upon its sign, the fermion zero-modes, which are 
identified with the known SM fermions, no longer have flat wave functions in 
the extra dimension. These are now found to be either peaked near $y=0$ or at 
the orbifold boundaries; this leads to potentially large differences in the 
various couplings of these fermions to the KK gauge fields from those expected 
in MUED. In particular, the zero mode fermions now have tree-level couplings 
to the KK-number even gauge modes. Second, the KK fermion wave functions and 
masses (which are given by $\sim n/R$ at tree-level) are now somewhat more 
complicated and are explicitly dependent upon the specific value 
of the $\mu$ parameter. In particular, the expressions for the KK fermion 
masses are {\it different} depending upon whether the relevant 
KK-number is even or odd.  

The purpose of this paper is ($i$) to explore in some of the detailed 
implications of non-zero values for fermion mass parameter $\mu$ leading 
to alterations from the conventional MUED 
phenomenology and ($ii$) to investigate the regions in the $R-\mu$ plane 
which are accessible to current and future collider experiments. 
To these ends, in Section \ref{sec:SUED}, we provide a basic overview of the masses, 
wave functions and couplings of the fermion KK states in 
split-UED model and display their 
explicit dependence upon the parameter $\mu$ pointing out important 
differences with the MUED case. Here we will assume that the $\mu$ parameter 
takes on a universal value for all fermions for simplicity of the analysis so 
that there is only one new parameter to consider beyond that of MUED. In 
Section \ref{sec:collider} we will discuss the collider phenomenology of split-UED 
and, in particular, the properties of the KK states and the potential for their 
discovery at the LHC. Further, we obtain the regions of the $R-\mu$ plane 
which are allowed by current experimental data and show the regions which will 
be made accessible by searches at the LHC. Our conclusions can be found in 
Section \ref{sec:conclusion}.
Appendix \ref{app:spectrum} contains detailed information of KK decomposition and  mass spectrum.

\section{Split Universal Extra Dimensions}
\label{sec:SUED}

\subsection{Model}
\label{sec:model}

Universal extra dimensions postulates that all of the Standard Model particles are propagating in a small extra dimension(s). Orbifold compactification makes it possible to construct a chiral four dimensional effective theory. 
In contrast to the brane world scenarios \cite{ADD,RS}, the translational symmetry along the extra dimension leads to a remnant discrete symmetry, dubbed KK parity, so that the lightest Kaluza-Klein particle can be a good dark matter candidate. Also this parity mimics the R-parity in supersymmetric theory so that UED phenomenology shares several common features with  MSSM \cite{Cheng:2002ab}. On the other hand, it has been often overlooked in UED models that the bulk Dirac masses are generically allowed and are not in conflict with higher dimensional 
Lorentz symmetry or gauge invariance. In this section, we review split-UED model where these bulk Dirac masses are generically allowed in a way that KK parity is intact. 

In split-UED, quarks $(Q, U^c, D^c)$ and leptons $(L, E^c)$ are all promoted to fields in five dimensional spacetime on $S^1/Z_2 \times M^4$ orbifold with two fixed points $y=- L$ and $y=L$, respectively, where $y$ is the coordinate along extra dimension with the half length $L=\pi R/2$. In the minimal setup, the gauge group is the same as in the Standard Model: $SU(3)_c\times SU(2)_W \times U(1)_Y$ under which charges are assigned as follows
\begin{eqnarray}
\Psi_i(x,y)=(Q_i, U^c_i, D^c_i, L_i, E^c_i)^c=((3,2)_{1/6},( \bar{3},1)_{-2/3},(\bar{3},1)_{1/6},(1,2)_{-1/2},(1,1)_{1})^c \, ,
\end{eqnarray}
where the index $i$ runs for three generations of fermions.
Allowing bulk mass term in split-UED 
the generic action $S=\int d^4 x \int_{-L}^L dy {\cal L}_5$ is given by
\begin{eqnarray}
{\cal L}_5= \sum_{i,j=1}^3 \frac{i}{2} (D_M \bar{\Psi}_i \Gamma^M \Psi_j -
\bar{\Psi}_i \Gamma^M D_M \Psi_j)  -m_{ij}(y) \bar{\Psi}_i \Psi_j \, ,
\label{Eq:action}\end{eqnarray}
where the covariant derivative is $D_M =\partial_M + i g_3 \frac{\lambda^\alpha}{2} G^\alpha_M + i g_2 \frac{T^a}{2} W^a_M + i g_1Y B_M $ with the usual Gell-Mann and Pauli matrices $\lambda$ and $T$. 
The $g_1$, $g_2$, $g_3$ and $G$, $W$, $B$ are the gauge coupling constants and 
the gauge fields of the corresponding gauge groups, respectively.
Without loss of generality we can diagonalize the action in Eq. (\ref{Eq:action}) 
by unitary transformations. Therefore the mass term $m_{ij}$ can be taken as 
$ m_{ij} = m_i \delta_{ij}$ and there is no kinetic mixing between different flavors (for $i\neq j$).
In general, we may have dimensionful parameters $(m_Q, m_{U^c}, m_{D^c}, m_L, m_{E^c})$ for each generation. Now imposing Dirichlet boundary conditions for unnecessary chiral component of fermions,
we can finally get exactly the same spectra in the SM for the lowest Kaluza-Klein modes. 
All the details of derivation to get the Kaluza-Klein spectra are described in the Appendix. 
The most prominent feature of split-UED is that the fermion profile in the extra dimension is either localized near the origin or at boundaries depending the sign of bulk mass parameter $m_i(y)=\mu_i \theta(y)$ in a way that Kaluza-Klein parity is respected. Having a non-zero bulk mass, $m$, a field still has a massless zero mode which satisfies Neumann boundary conditions. However its Kaluza-Klein excitation states get additional contributions and the mass is given by 
$m_n = \sqrt{k_n^2 + \mu^2}$ where $k_n$ is the momentum to the extra dimension which is determined by $\mu= \pm k_n \cot k_n L$ for $n\in Z_{\rm odd}$ or $k_n = \pi n/L$ for $n\in Z_{\rm even}$. Here we impose Dirichlet boundary conditions for $\Psi_L$ modes so that $\Psi_R$ contains the SM fermions in our 
convention. We assume that gauge sector and Higgs sector remain the same as in the conventional UED models. Therefore the zero modes have flat profiles and Kaluza-Klein modes have cosine wave functions satisfying Neumann boundary conditions.

In summary, in split-UED, there are new $15$ dimensionful parameters $\mu_\Psi$ for three generations, 
the cutoff scale ($\Lambda$) and one length parameter $L$ ($=\pi R/2$) given by the size of extra dimension in addition to the SM parameters.
In this study we will consider all bulk mass parameters are the same for simplicity, 
and study $\mu L \geq -1$ region
\footnote{For $\mu L < -1$, the KK spectra contain the unacceptable light modes 
below KK scale $\sim$ TeV.}.

\subsection{Behavior of couplings}
\label{sec:couplings}

Having the explicit wave functions of fields as given in the Appendix, we can calculate the explicit Lagrangian for interactions among those fields. Essentially the overlap integral of the wave functions gives the effective couplings. For a gauge boson $V=(G, W, B)$ after choosing a simplifying gauge to get rid of the fifth component of 
gauge multiplet, $V_5$, by the orbifold condition, we find 

\begin{eqnarray}
-{\cal L}_{\rm int} &\ni&  g_V \int_{-L}^L d y  \, \bar{\Psi}\Gamma^\mu \Psi V_\mu \\
&=& g_V \sum_{\ell mn}\int_{-L}^L d y \, 
\Big [ \bar{\psi}^\ell {f^\ell_\Psi}^* (y) \Big ] \gamma^\mu
\Big [ \psi^m f_\Psi^m(y) \Big ]
\Big [ V_\mu^n f_V^n(y) \Big ] \\
&=& \sum_{\ell mn} g^{eff}_{\ell mn} \bar{\psi}^\ell \gamma_\mu \psi^m V_\mu^n \, ,
\end{eqnarray}
where the effective coupling is obtained by the integration of the wave function overlap
with a convenient dimensionless variable $x_\Psi=\mu_\Psi L$:
\begin{eqnarray}
g^{eff}_{\ell mn} &&\equiv g_V \int_{-L}^L  d y \, {f^\ell _\Psi}^*(y) f_\Psi^m(y) f_V^n(y) \\
&&\equiv g_V {\cal  F}_{\ell mn}(x_\Psi) \, .
\end{eqnarray}
As the profiles of gauge bosons are universal and the profile of fermions depend on 
the bulk mass parameter $\mu_\Psi$, the overlap integral ${\cal F}_{\ell mn}$ is the same for all gauge bosons 
but depends on $\mu_\Psi$. The suppressed gauge group indices should be understood.

Let us now see  the coupling between KK bosons ($G, W, B$) of $SU(3)_c$, $SU(2)_W$ and $U(1)_Y$ and the zero mode SM fermion pair for the definiteness \footnote{As it is clear in minimal UED, the weak mixing angles for KK gauge bosons are suppressed by $m_W/m_{\rm KK} \ll 1$. Thus essentially gauge eigenstates are well aligned by mass eigenstates.}. 
The zero mode wave function profile for SM fermion $\Psi_i=(Q,U^c, D^c, L,E^c)^c$ is given by 
\begin{eqnarray}
f_i^{(0)}(y) = \sqrt{\frac{\mu_i}{1-e^{-2\mu_i L}}} e^{-\mu_i |y|} \, .
\label{Eq:zeromode}
\end{eqnarray}
If $\mu_i>0 (<0)$, the profile is exponentially localized near the center (at the boundaries). The zero mode is massless in the absence of the electroweak symmetry breaking even though its KK modes get additional mass from the bulk mass $\mu_i$. 

\FIGURE[t]{
\centerline{
\epsfig{file=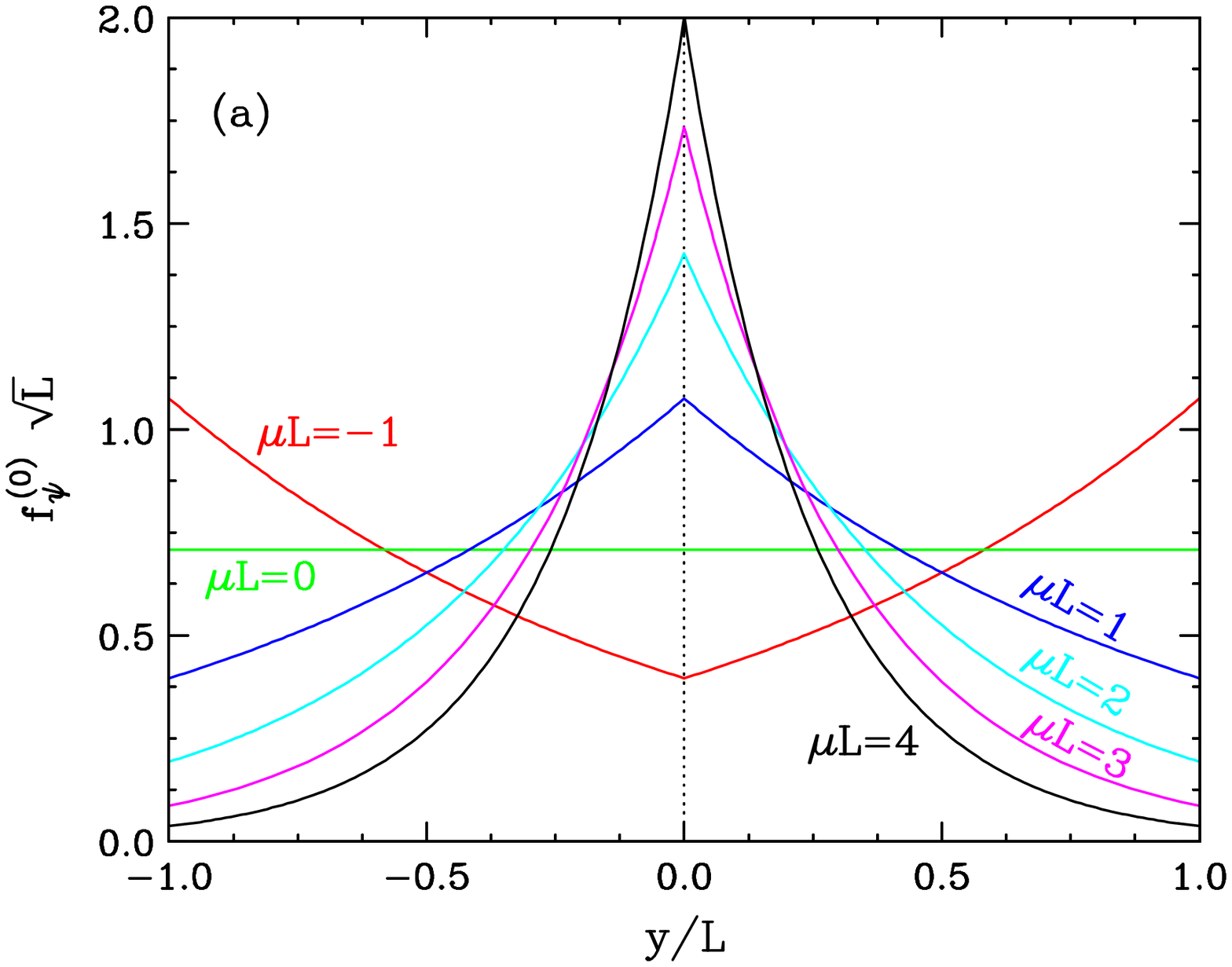, width=7.6cm} \hspace{0.0cm} 
\epsfig{file=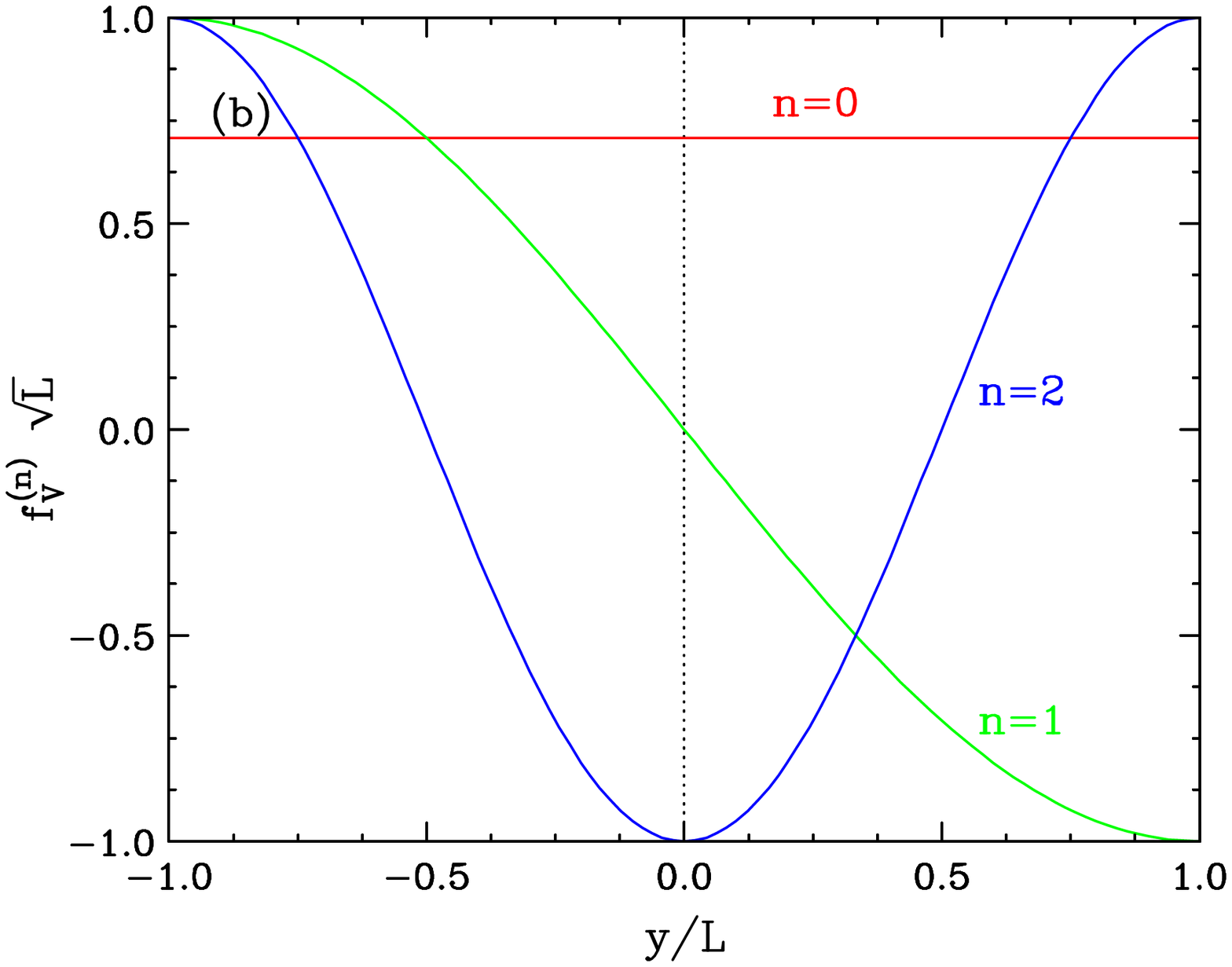, width=7.6cm}}
\caption{\sl Profiles of zero mode fermion with various $\mu$'s in (a)
and first three KK gauge bosons $n=0,1,2$ in (b). Kaluza-Klein parity
is obviously respected by the zero mode profile and the localization depends on 
the sign of the bulk mass $\mu$. KK-odd mode ($n=1, 3, 5, \cdots$) are odd  and even modes ($n=0,2,\cdots$) are even under KK parity. 
}
\label{fig:profile}}

The KK gauge bosons  commonly have the same profiles as in MUED 
\begin{eqnarray}
f_{V=G, W, B}^{(n>0)}(y)=\frac{1}{\sqrt{L}} \cos \frac{ n \pi (y+L) }{2L} \, ,
\label{eq:vec}
\end{eqnarray}
and the zero mode profile, $f_V^{(0)}=1/\sqrt{2L}$, is flat, as shown in Fig.~\ref{fig:profile}. 
Note that $\int_{-L}^L dy \, \big (f_V^{(n)} \big )^2=1$.
The coupling of level-$n$  bosons to SM fermion pair is now written as
\begin{equation}
{\cal L}_{\rm eff}\ni -\sum_\psi  \sum_n  \frac{C_{n}(\mu_\psi)}{\sqrt{2L}} \left[\bar{\psi_0}\gamma^\mu \left(g^{5D}_3\frac{\lambda^a}{2} G^{a,(n)}_\mu +g^{5D}_2\frac{T^i_\psi}{2}  W_\mu^{i,(n)} + g^{5D}_1 Y_\psi B_\mu^{(n)}\right)
\psi_0\right] \, , \, 
\label{Eq:int}
\end{equation}
where $C_n$ is a dimensionless  parameter measuring the overlap of wave functions 
between two SM fermions and a KK gauge boson defined as
\begin{eqnarray}
C_{n}(\mu_\psi L)&\equiv& \sqrt{2L}\int_{-L}^{L} dy (f_\psi^{(0)})^2 f_V^{(n)} \\
&=& \left\{ \begin{array}{ll}
0 ,                     & \hspace{1cm}\mbox{$n =1, 3, 5, 7,  \cdots$};\\
{\cal F}_{00n}(x_\psi) , & \hspace{1cm}\mbox{$n = 0, 2, 4, 6, 8, \cdots$} ,\end{array} \right. 
\end{eqnarray}
where $x_\psi = \mu_\psi L$ and ${\cal F}$ is explicitly calculated to be 
\begin{equation}
{\cal F}_{002m} (x) 
=\frac{x^2 (-1+(-1)^m e^{2x})(\coth x-1)}{\sqrt{2(1+\delta_{m0})}(x^2 + m^2\pi^2/4)} \, ,
~~~ m=0, \, 1, \, 2, \,3 , \, \cdots \,  .
\end{equation}
From the KK-parity conservation, $C_{odd}=0$ is easily understood. The Standard Model coupling constants are obtained as 
\begin{eqnarray}
g^{SM}=g_{000}^{eff}=\frac{g^{5D}}{\sqrt{2L}} C_0 (x_\psi) = \frac{g^{5D}}{\sqrt{2L}} \, ,
\label{eq:gsm}
\end{eqnarray}
as $C_0(x_\psi)={\cal F}_{000}(x_\psi)=1$ independent of $x_\psi$. Here $g_{\ell nm}^{eff}$ 
denotes the effective coupling constant for the $\psi_\ell -\psi_m-V_n$ interaction.
Finally for the even $n$'s we get the coupling between the Standard Model fermions and even-KK excitation states of gauge bosons, as
\begin{eqnarray}
g_{002n}^{eff}=g^{SM} {\cal F}_{002n} (x_\psi) \, .
\end{eqnarray}

\FIGURE[t]{
\centerline{
\epsfig{file=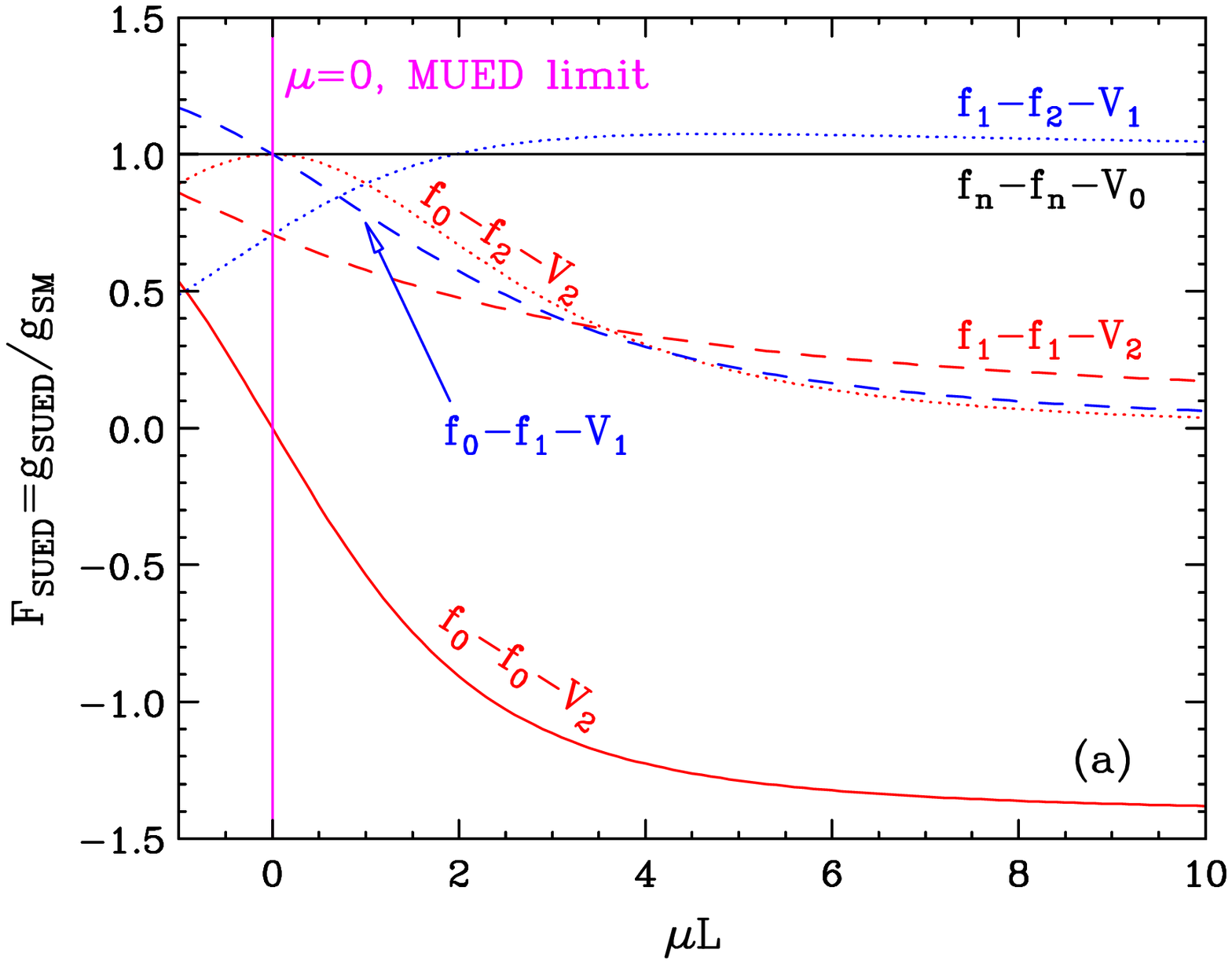, width=7.5cm}  
\epsfig{file=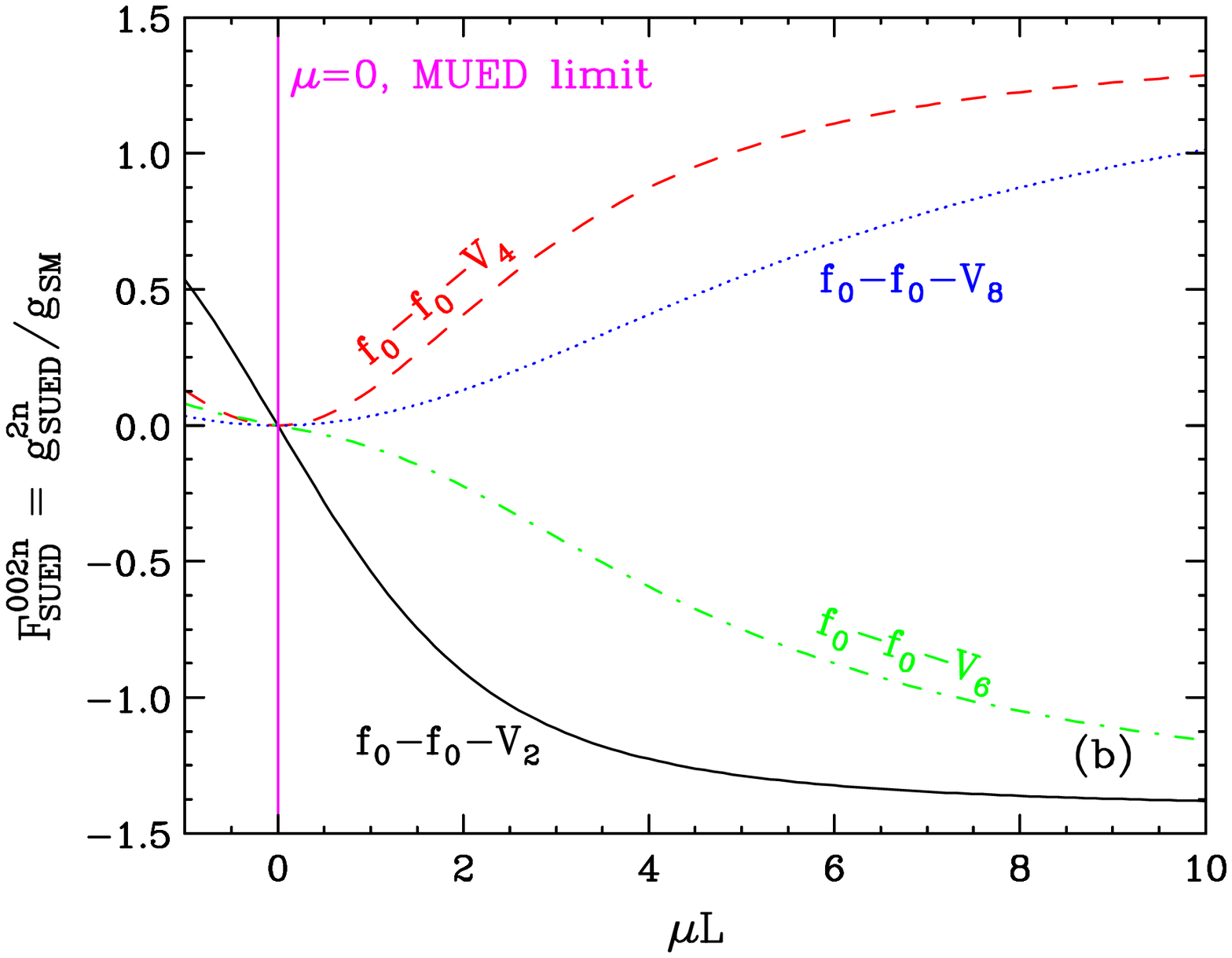, width=7.5cm}}
\caption{\sl The ratio of tree level couplings in SUED to the corresponding SM couplings. 
Couplings involving level-2 (level-1) KK bosons are shown in red (blue) in (a). (b) contains zero mode fermion couplings to KK-even gauge bosons $f_0-f_0-V_{2n}$. 
The MUED limit ($\mu=0$) is denoted by the vertical solid line (in magenta).
}
\label{fig:couplings}}
The various couplings associated with one vector boson and two fermions are 
shown in Fig.~\ref{fig:couplings}.
We find that there are two interesting regions.
One is the MUED limit, {\it i.e.,} $\mu \to 0$, which is 
shown as the vertical solid line (in magenta) and the other is large positive $\mu$ limit.
For $\mu \to \infty$, the zero mode fermions are well localized near the center ($y=0$) 
so that their couplings to KK gauge bosons asymptotically approach to the well known value $(-1)^n \sqrt{2}$ as one can see the red 
curve for $f_0-f_0-V_2$ (Fig.~\ref{fig:couplings}(a)) 
as well as the curves for $f_0-f_0-V_{2n}$ (Fig.~\ref{fig:couplings}(b)). 
The alternating sign can be understood as the $2n$-th KK gauge boson wave function in Eq. (\ref{eq:vec}) is proportional to $\cos n\pi = (-1)^n $ at $y=0$ where the fermion wave function is mostly localized. The $\sqrt{2}$ is from the zero mode normalization in Eq. (\ref{eq:gsm}). These vertices are all vanishing in the limit of $\mu\to 0$ because of KK number conservation in MUED.

For collider phenomenology, we are mostly interested in interactions for 
low lying Kaluza-Klein modes, $n=0,1,2$ as heavier modes are too massive and 
easily decouple from the low energy phenomenology. 
The most relevant couplings in our study are the interactions 
and decays of the second Kaluza-Klein gauge boson.

The coupling $f_n$-$f_n$-$V_0$ remains the same for all $\mu$ due to the normalization condition of 
wave functions for $n$-th fermion profile, while all other couplings now change 
for non-vanishing bulk masses.
The $f_2$-$f_0$-$V_0$ coupling remains zero in SUED 
but in principle this coupling can be generated 
by the unknown physics at the cutoff scale ($\Lambda$), and the lowest order coupling may take 
the form \cite{Cheng:2002iz}
\begin{equation}
\bar{f}_2 \sigma^{\mu\nu} T^a P_{L/R} f_0 F_{0\mu\nu}^a \, .
\end{equation}
However, being higher dimensional, we expect it to be suppressed at least 
by one power of $1/\Lambda$, hence we shall neglect it in the discussion that follows.

It is interesting to notice that the $SU(3)_c$ coupling for the KK gluon can be {\it chiral}. 
Let us examine the level-$2n$ gluon couplings with quarks:
\begin{eqnarray}
-{\cal L}_{\rm eff}=g_s\sum_{n \geq 0} [
\bar{u}\gamma^\mu \left({\cal F}_{002n}(x_Q)P_L + {\cal F}_{002n}(x_U)P_R \right)u\nonumber \\
+\bar{d}\gamma^\mu \left({\cal F}_{002n}(x_Q)P_L + {\cal F}_{002n}(x_D)P_R \right)d]G_\mu^{(2n)} \, .
\end{eqnarray}
All the KK-parity violating interactions are forbidden. 
Now it is obvious that KK-gluon has chiral interactions with the SM quarks, 
if $\mu_Q \neq \mu_U$ or $\mu_Q \neq \mu_D$, in general. 
Finally the vector (V) and axial-vector (A) couplings of KK gluons 
with an up-type quark and a down-type quark, 
\begin{eqnarray}
-{\cal L}_{\rm eff}=g_s \sum_{q=u,d}\sum_{n\geq 0}\bar{q}\gamma^\mu (V^q_{2n} -A^q_{2n}\gamma_5) q G_\mu^{(2n)} \, ,
\end{eqnarray}
are determined as
\begin{eqnarray}
V_{2n}^{u/d} =\frac{1}{2}\left({\cal F}_{002n}(x_Q)+{\cal F}_{002n}(x_{U/D})\right) \\
A_{2n}^{u/d} =\frac{1}{2}\left({\cal F}_{002n}(x_Q)-{\cal F}_{002n}(x_{U/D})\right) \, .
\end{eqnarray}
The same is similarly true for all other gauge bosons as well. When $x_Q=x_{U/D}$, only the vectorial coupling is non-vanishing.  However, in general, $x_Q \neq x_{U/D}$ and the non-vanishing axial couplings are allowed. For instance, if $x_Q=0$ and $x_{U/D}\neq 0$, the vectorial and axial couplings have opposite signs but have the same size: $V_{2n}^{u/d}=-A_{2n}^{u/d}={\cal F}_{002n}(x_{U/D})$. With non-vanishing axial couplings in the even KK gauge boson interactions, one might expect, for instance, additional contribution to the forward-backward asymmetry of the top quark pair production ($A_{FB}^t$) via the quark pair annihilation channel. The cross-section for $q\bar{q}$ annihilation into top quarks of mass $m_t$ 
through the $2n$-th KK gluons reads   
\begin{eqnarray}
\frac{d\sigma (q\bar{q}\rightarrow g_{2n}^* \to t \bar{t})}{d\cos \hat{\theta}} &=&
\frac{\pi \beta \alpha_S^2}{9 \hat{s}} \Big \{ 1+c^2 \beta^2+ \frac{4 m_t^2}{\hat s}    \nonumber  \\ 
 && \hspace*{-3cm} + 
\sum_{n \geq 1} \frac{2 \hat{s} (\hat{s}-m_{2n}^2)} {(\hat{s}-m_{2n}^2)^2+m_{2n}^2 \Gamma_{2n}^2}
\Big [ V_{2n}^q \, V_{2n}^t \, \big (1+c^2 \beta^2+ \frac{4 m_t^2}{\hat s} \big )    
+ 2 \, A_{2n}^q \, A_{2n}^t \, c \beta \Big ]   \nonumber    \\
&&  \hspace*{-3cm}+
\sum_{n,\ell \geq  1} \hat{s}^2 \frac{(\hat{s}-m_{2n}^2)(\hat{s}-m_{2\ell}^2) + m_{2n} m_{2\ell} \Gamma_{2n} \Gamma_{2\ell}} 
       { [ (\hat{s}-m_{2n}^2)^2+m_{2n}^2 \Gamma_{2n}^2 ] [ (\hat{s}-m_{2\ell}^2)^2+m_{2\ell}^2 \Gamma_{2\ell}^2 ] } \\
&& \hspace{-2.5cm}\times  \Big [  \Big ( V_{2n}^q V_{2\ell}^q + A_{2n}^q A_{2\ell}^q \Big )
\Big ( V_{2n}^t V_{2\ell}^t  \big (1+c^2 \beta^2+ \frac{4 m_t^2}{\hat s} \big ) 
+   A_{2n}^t A_{2\ell}^t \beta^2 (1+c^2) \Big )      \nonumber \\
   && \hspace{-1cm} + 2 \, c \beta \, \Big ( V_{2n}^q A_{2\ell}^q + V_{2\ell}^q A_{2n}^q \Big )
                          \Big ( V_{2n}^t A_{2\ell}^t + V_{2\ell}^t A_{2n}^t \Big )  \Big ]     \nonumber   
\Big\} \, ,    \nonumber 
\label{eq:bornqq}
\end{eqnarray}
where $\hat{\theta}$ is the polar angle of the top quark with respect
to the incoming quark in the center of mass rest frame,
$\hat{s}$ is the squared partonic invariant mass,
$\beta = \sqrt{1-\frac{4 m_t^2}{\hat s}}$ is the velocity of the top quark,
with $c = \cos \hat{\theta}$.
The parameters $V_{2n}^q (V_{2n}^t)$ and $A_{2n}^q(A_{2n}^t)$ represent, 
respectively, the vector and axial-vector couplings of the
KK gluons to the light quarks (top quarks). 

Considering experiments at Tevatron, the parton level energy $\hat{s}$ is typically 
much less than the KK gluon mass so that the interference term (the second term) is 
dominant over the pure new physics term (the third term).
And the leading contribution in the second term is 
the interference between two diagrams the SM gluon and the level-2 KK gluon.
As the tree level SM contribution (the first term) does not produce 
the forward-backward asymmetry after integrating over 
$-1< \cos \hat{\theta}<1$, the main contribution is from the linear term of cosine 
in the second term for $n=1$:
\begin{eqnarray}
A_{FB}^t 
&\propto & -  \frac{A_{2}^q A_{2}^t}{m_{2}^2}.
\end{eqnarray}
When $x_t = -1$, $x_{U/D} \to \infty$ and $x_Q=0$, 
\begin{eqnarray}
A_{2}^q \to \frac{1}{\sqrt{2}}, \,\, A_{2}^t \to -\frac{1}{4},
\end{eqnarray}
thus the forward-backward asymmetry is positive, which is consistent with the recent measurements at Tevatron 
\cite{newcdf, cdf, d0}, but we find that its size is not large enough to explain the current anomaly 
for $R^{-1} \sim 1$ TeV.

\subsection{Mass spectrum}
\label{sec:masses}

The mass spectrum of fermions gets tree level modifications from the bulk parameters $\mu_\Psi$ as well as the loop induced mass correction from the RG running effect for a given boundary condition at some high scale $\Lambda$ just as in the case in conventional UED. Taking the vanishing boundary condition at $\Lambda$, it is known that the one-loop induced mass correction is minor ($\sim \%$ level for electroweak particles). This is due to lack of long RG running from $\Lambda$ which is argued to be less than $100$ TeV based on naive dimensional analysis (see e.g. \cite{Cacciapaglia:2005da}). Thus we may neglect the loop-induced mass correction for fermions as long as the bulk mass parameter is sufficiently large $\mu_\Psi > 0.1/L$.

The mass of KK fermon ($M_n$) gets contributions from the bulk mass at tree level 
as follows
\begin{eqnarray}
M^2_{n} = k_n^2 + \mu^2 \, ~~~ {~\rm for ~} n\ge 1 \, ,
\end{eqnarray}
where
\begin{eqnarray}
&& k_n {\rm ~is~the~}\frac{n+1}{2}-{\rm th~solution~of~} \mu=-k \, cot(k L) \, , 
                         {\rm ~if~} n=2m-1 \, ,\label{eq:oddmodes}\\
&& k_n = \frac{n}{R}\, , \hspace*{6.75cm}{\rm ~if~} n=2m \, .   \label{eq:evenmodes}
\end{eqnarray}
In the MUED limit, $\mu \to 0$, Eqs.~(\ref{eq:oddmodes}-\ref{eq:evenmodes}) both 
reduce to $k_n = \frac{n}{R}$. On the other hand, all KK boson masses remain the same, 
$\frac{n}{R}$, and show no $\mu$ dependence. 

Including EW symmetry breaking and the radiative corrections, a naive estimate gives 
\begin{eqnarray}
M_n &\approx& M_n^{tree} \left ( 1 + {\rm ~ loop ~corrections}   \right ) \, , \\
M_n^{tree} &=& \sqrt{ k_n^2 + \mu^2 + m_0^2} \, ,
\end{eqnarray}
where $m_0$ is expected from the electroweak symmetry breaking.

\subsection{Constraints from contact interactions}
\label{sec:constraints}

One of the most prominent features of SUED having non-vanishing bulk mass parameters is 
the existence of tree level KK number violating interactions. 
From $W^3_{2n},B_{2n}$ exchange diagrams we can effectively obtain the contact interaction Lagrangian ${\cal L}_{\rm eff}$  which is stringently constrained by electroweak precision measurements 
\cite{Alcaraz:2006mx,Amsler:2008zzb}
\begin{eqnarray}
{\cal L}_{\rm eff}=\sum_{i,j=L,R}\sum_{f}\frac{4\pi}{(\Lambda_{AB}^{ef})^2}
[\bar{e_i}\gamma_\mu e_i][ \bar{f}_j \gamma^\mu f_j] \, .
\label{Eq:eff}
\end{eqnarray}
%
%
\TABLE[t]{
\centerline{
\begin{tabular}{c|c|c|c|c}
{}&u&d&$\mu^+\mu^-$ &$\tau^+\tau^-$\\
\hline
$LL$ (TeV) &  10.2  &  6.0  &  12.5  &  8.6 \\
$RR$ (TeV) &  8.3   &  4.3  &  11.9  &  8.2
\end{tabular}
}
\label{table:contact}
\caption{Bounds for contact interaction \cite{Alcaraz:2006mx,Amsler:2008zzb}}.}
Assuming a universal bulk mass $\mu$, the $B_{2n}$ and $W^3_{2n}$ mediated interaction 
effective Lagrangian is obtained. 
The most stringent bound comes from the contact interaction for $ee\mu\mu$:
\begin{equation}
\bar{e}_L \gamma_\mu e_L \sum_{n} \frac{({\cal F}_{002n})^2}{4}
\left(\frac{g_1^2}{m_{B_{2n}}^2} +\frac{g_2^2}{m_{W^3_{2n}}^2}\right)\bar{\mu}_L \gamma^\mu \mu_L  
+
\bar{e}_R \gamma_\mu e_R \sum_{n} ({\cal F}_{002n})^2
\left(\frac{g_1^2}{m_{B_{2n}}^2}\right)\bar{\mu}_R \gamma^\mu \mu_R \, .
\label{Eq:eff2}
\end{equation}

Taking Eqs. (\ref{Eq:eff}-\ref{Eq:eff2}) into account with $m_{B_{2n}}\simeq m_{W^3_{2n}}\simeq (2n)/R$, we obtain the following relations:
\begin{eqnarray}
\frac{1}{\Lambda_{LL}^2}&=&\frac{g_1^2+g_2^2}{64\pi} R^2 \sum_n \frac{({\cal F}_{002n}(\mu L))^2}{n^2},
\\
\frac{1}{\Lambda_{RR}^2}&=&\frac{g_1^2}{16\pi} R^2 \sum_n \frac{({\cal F}_{00,2n}(\mu L))^2}{n^2},
\end{eqnarray}
where the bounds for $\Lambda_{LL}$ and $\Lambda_{RR}$  are given in Table \ref{table:contact}. 

We also consider the constraints arising from the dilepton resonance searches at Tevatron \cite{CDFdimuon}
and find that those for $\gamma_2$ give a slightly better constraint on $R^{-1}$ 
than those for $Z_2$, while $W_2^\pm$ gives a similar limit to that for $Z_2$ \cite{:2007bs}. 
In the next Section we include these as well as constraints from contact interactions.

\section{Collider phenomenology}
\label{sec:collider}

A large amount of effort has given into examining the collider aspects of 
Universal Extra Dimensions \cite{Appelquist:2000nn} 
at LHC \cite{Rizzo:2001sd,Cheng:2002ab,Datta:2005zs,Burdman:2006gy,Dobrescu:2007xf} and 
ILC \cite{Battaglia:2005zf,Freitas:2007rh}, 
as well as its astrophysical implications \cite{Servant:2002aq,Cheng:2002ej,Kong:2005hn,Burnell:2005hm,Dobrescu:2007ec,Arrenberg:2008wy}.
In this Section we would like to investigate the implications of non-vanishing bulk mass in SUED.

\subsection{Level-1 modes}
\label{sec:level1}

We start our discussion with the level-1 KK modes.
Their phenomenology depends on the precise value of the bulk mass and 
the radiative corrections to KK masses. Therefore here we would like discuss only 
generic features.

A small value for the bulk mass ($0 \leq|\mu L| \ll 1$) would give the similar decay patterns 
as in the MUED case. The dominant production is provided by the strong interaction 
at a hadron collider, {\it i.e.,} KK quark production ($Q_1Q_1$, $q_1q_1$ and $Q_1 q1$), 
KK gluon production ($g_1g_1$) and associated production ($g_1Q_1$, $g_1q_1$).
The $SU(2)_W$-doublet KK quarks ($Q_1$) dominantly decay into 
$SU(2)_W$ KK gauge bosons ($Z_1$ and $W_1^\pm$) while 
the $SU(2)_W$-singlet KK quarks ($q_1$) decay into KK photon.
SM leptons are obtained from the decay of EW gauge bosons ($Z_1$ and $W_1^\pm$) 
to KK leptons.
Here the difference with MUED would be the mass splitting between each mode.
A bulk mass term would increase the mass of the KK fermion, 
making the decay product of KK bosons softer than that in MUED.
However the other decay products from the KK fermions to KK bosons become more energetic 
due to the increased splitting. 
For instance, KK quarks can have mass just below the KK gluon and 
the jet from the decay of KK gluon ($g_1 \to Q_1 q$) would be softer 
while the jet from the decay of KK quark ($Q_1 \to q Z_1$ or $Q_1 \to q' W_1^\pm$) 
becomes harder than in MUED. The same is true for KK leptons and KK gauge bosons.
However we do not expect a dramatic change in the reach 
for this model, as long as the decay patterns are the same and the mass splitting is not too small.

The other extreme limit is the case of very large $\mu$, $\mu L \gg 1$
\footnote{The opposite limit (a large negative $\mu L$) 
is also interesting as shown in the Appendix.
In this case, taking a very large $R^{-1}$, the masses of the level-1 KK fermions remain at EW scale 
while all other KK fermions at higher level are decoupled from the theory due to the large mass splitting 
between the level-1 and the level-2 KK fermions. All KK bosons are also very heavy due to the large $R^{-1}$.
Therefore in this limit, {\it the only} available KK modes are the level-1 KK leptons and quarks. 
This study will be elsewhere \cite{ongoing}.}.
In this case, all KK fermions become much heavier than KK bosons and 
they may not be within the reach of the LHC. KK gauge bosons go through 3 body decays 
to the KK photon ($g_1 \to j j \gamma_1$ and $Z_1,W_1^\pm \to f\bar{f}' \gamma_1$) and 
production would be via the KK gluon ($g_1g_1$) and EW gauge bosons 
($Z_1 W_1^\pm$) 
\footnote{$Z_1Z_1$, $\gamma_1\gamma_1$ and $Z_1\gamma_1$ involve KK fermions 
in the $t$- and $u$-channels and their production cross sections are negligible 
for heavy KK quarks.}.
It is interesting to notice that this situation is similar to 
{\it the focus point region of supersymmetry}. 

For the moderate ranges of the bulk mass ($|\mu L| \sim 1$), 
the gauge bosons may still go through 3-body decays 
while the LHC will be able to produce KK quarks and KK leptons.
Unlike MUED, now all KK quarks dominantly decay to the KK gluon since it has the largest coupling.
Therefore the collider signature would be quite jetty.

An interesting possibility is that all KK fermions are very heavy with a large $\mu$ 
so that they are unobservable but still the KK bosons are within the reach of the LHC. Even in this case, we might expect to observe e.g. the level-2 gluon through its interaction with the level zero Standard Model quarks with a sizable coupling ($g\simeq \sqrt{2}g_s$). Dilepton production through $Z_2$ and $\gamma_2$ is also sizable and provides a golden search channel for split-UED. Detection of a dark matter (DM) particle, on the other hand, is quite challenging because the DM-SM coupling through the level-1 fermion will be highly suppressed by the large KK fermion mass.

We plot cross sections for the gauge boson production at the LHC, as function of mass 
in Fig.~\ref{fig:xsection1}, assuming $\mu L \gg 1$; here the curve is 
dotted for $g_1 g_1$ (in black), solid for $Z_1W_1^+$ (in red) and dashed for $Z_1W_1^-$ (in blue).
For KK gluon pair production from the $gg$ initial state, 
there are $s$-, $t$-, $u$- and four point interaction diagrams, and 
all couplings are fixed by $SU(3)_c$ gauge invariance. 
There is a contribution from the $q\bar{q}$ initial state but it is smaller than that from $gg$ at the LHC 
for the mass range shown.
$Z_1 W_1^\pm$ is produced by $W^\pm$ exchange in the $s$-channel and 
contribution from KK quarks in the $t$- and $u$-channels. 
Having these 3 diagrams is also necessary by $SU(2)_W$ gauge invariance and 
neglecting any of them gives inconsistent results. 
However in our case, considering the limit where the KK quarks are much heavier than KK gauge bosons, 
their exchanges with a few TeV in mass barely affects the production cross sections.

\FIGURE[t]{
\centerline{
\epsfig{file=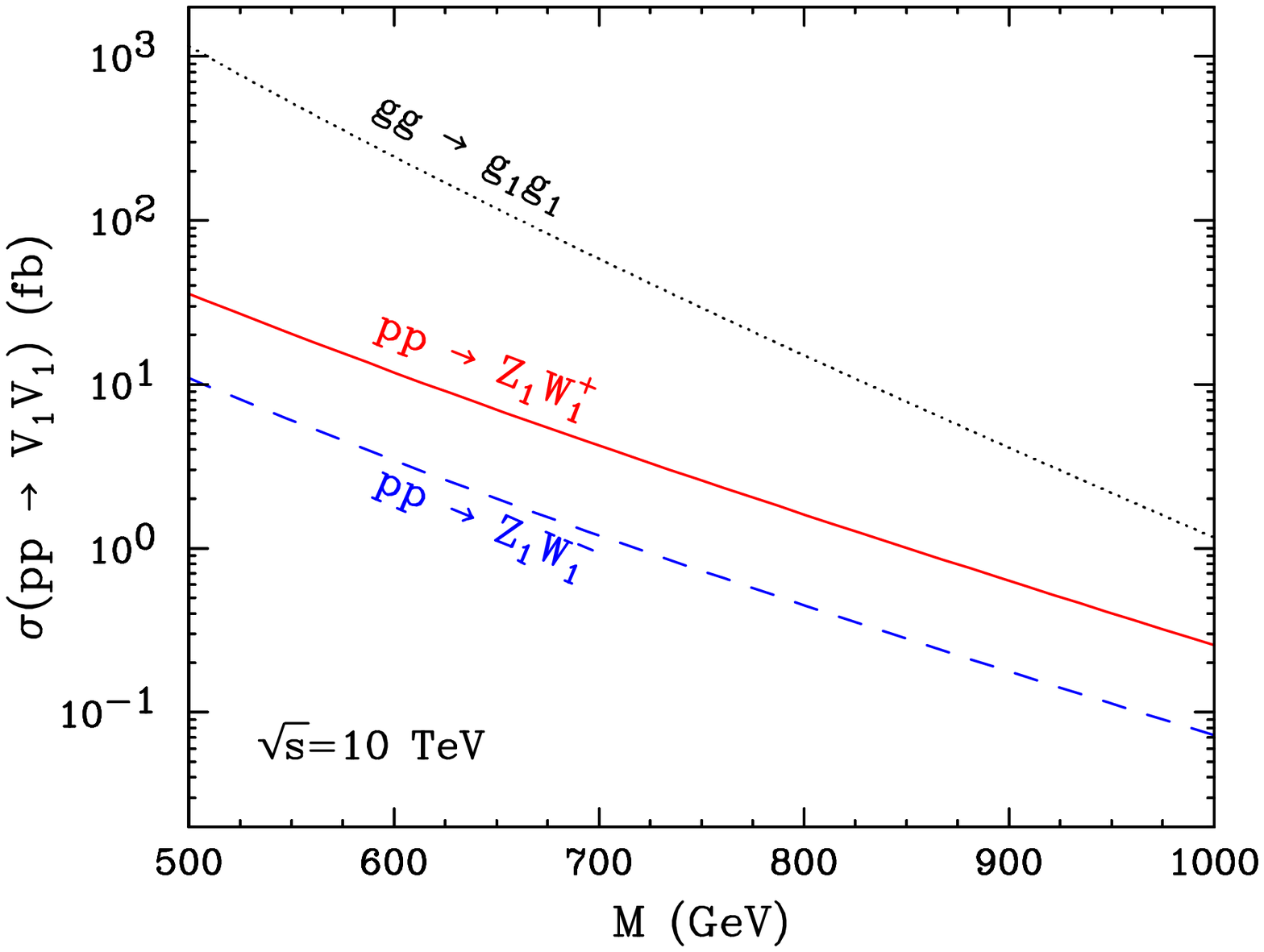, width=8.5cm}  }
\caption{\sl Cross section for gauge boson pair production as a function of mass at the LHC (for $\mu L \gg 1$).}
\label{fig:xsection1}}

From Fig.~\ref{fig:couplings} we can immediately read off the patterns of KK particle decay branching fractions. 
The level-1 gauge boson only couples to $f_0-f_1$ thanks to KK parity. This coupling becomes less significant as $\mu$ becomes larger in which case the level-1 fermion becomes significantly heavier than 
the level-1 gauge boson. Due to the large mass gap, the decay products of the level-1 fermion are 
reasonably energetic.

\subsection{Level-2 modes}
\label{sec:level2}

Now we turn to discussion of the level-2 KK modes.
In general level-2 KK fermions ($f_2$) can decay into 
either two level-1 KK states $f_1V_1$
or one level-2 and one SM mode $f_0 V_2$ 
(the branching fraction of $f_2$ to $f_2^\prime V_0$ is suppressed by the small mass splitting between 
$f_2$ and $f_2^\prime$.).  
In the limit of $\mu L \gg 1$ in split-UED, all KK fermions masses are raised, and 
level-2 KK quarks ($Q_2$ and $q_2$) decay to $q g_2 $ and $Q_1g_1$.
Then $g_1$ gives rise to a missing energy signature through a 3 body decay 
while $g_2$ can directly decay into two jets and may appear as dijet resonance.

In MUED, the coupling of level-2 resonances to the SM fermions is suppressed by 1-loop. 
The branching fractions of electroweak level-2 gauge bosons 
into dilepton final states are small partly due to the competing decay modes into 
other level-2 and level-1 KK states, and partly due to difference between 
the strong and electroweak couplings.
Therefore one has to rely on indirect production of level-2 KK gauge bosons from 
the KK gluon and KK quarks to enhance the production cross sections.
The corresponding reach has been estimated in Ref. \cite{Datta:2005zs}.

In SUED, however, this coupling exists at tree level 
due to the fermion bulk mass term and 
it could be as large as $\sqrt{2}$ times the corresponding SM coupling strength, which makes 
dilepton searches in this model promising.
At the same time, the bulk mass increases the mass of KK fermions, 
thus reducing the branching fraction of level-2 bosons into other KK states.
The decay width of level-2 KK bosons into SM fermion final states is given by
\begin{eqnarray}
\label{eq:width1} \Gamma &=& \frac{N_c M}{24 \pi} \left [ 
\Big ( g_L^2 + g_R^2 \Big ) \Big ( 1- \frac{m^2}{M^2} \Big ) 
+ 6 g_L g_R \frac{m^2}{M^2} \right ] \sqrt{1 - \frac{4m^2}{M^2}} \\
&=& \label{eq:width2} 
\frac{N_c M }{ 24 \pi} \Big ( g_L^2 + g_R^2 \Big )    ~~~~{\rm for ~ M \gg m} \, .
\end{eqnarray}
\FIGURE[t]{
\centerline{
\epsfig{file=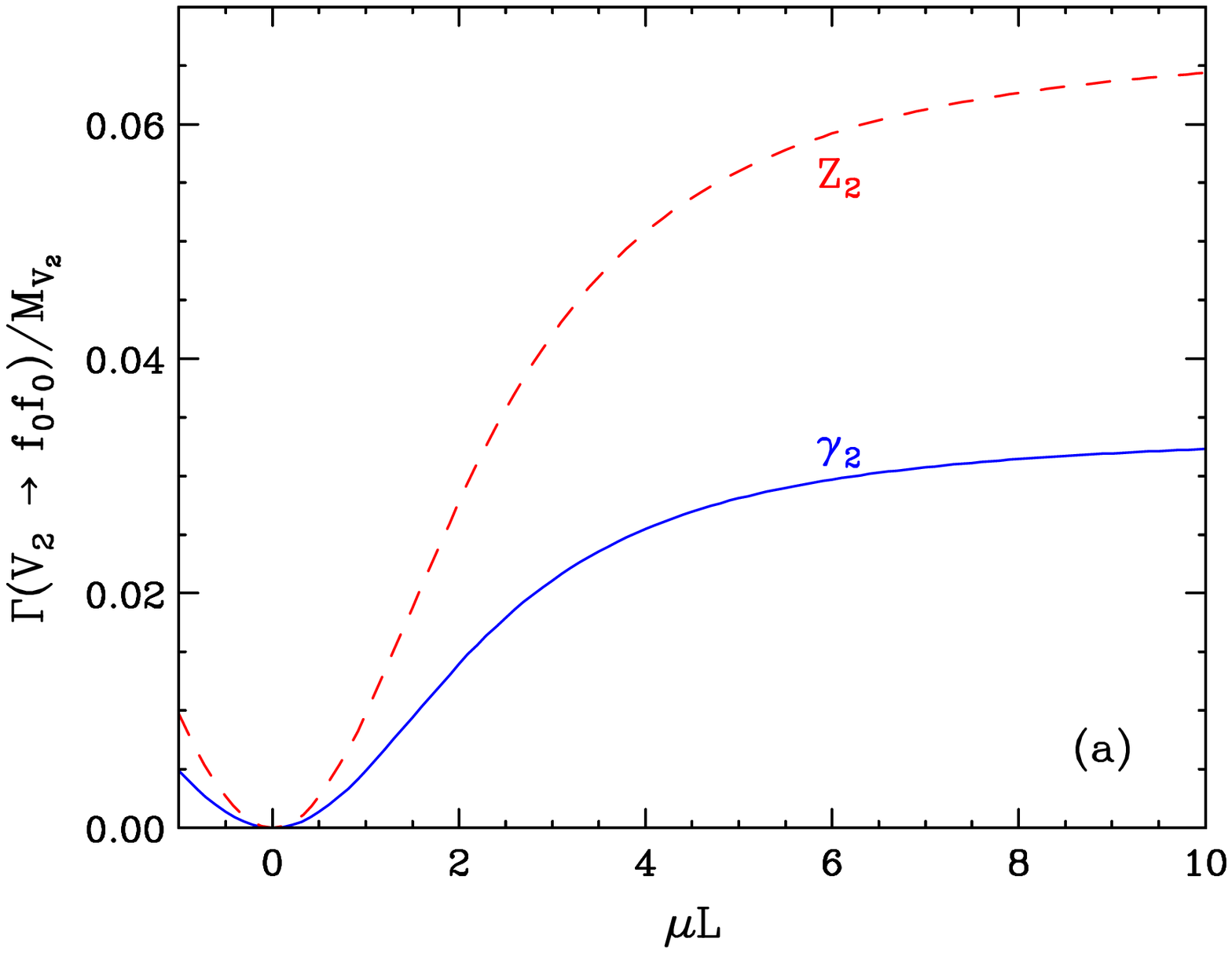, width=7.5cm} \hspace*{0.1cm}
\epsfig{file=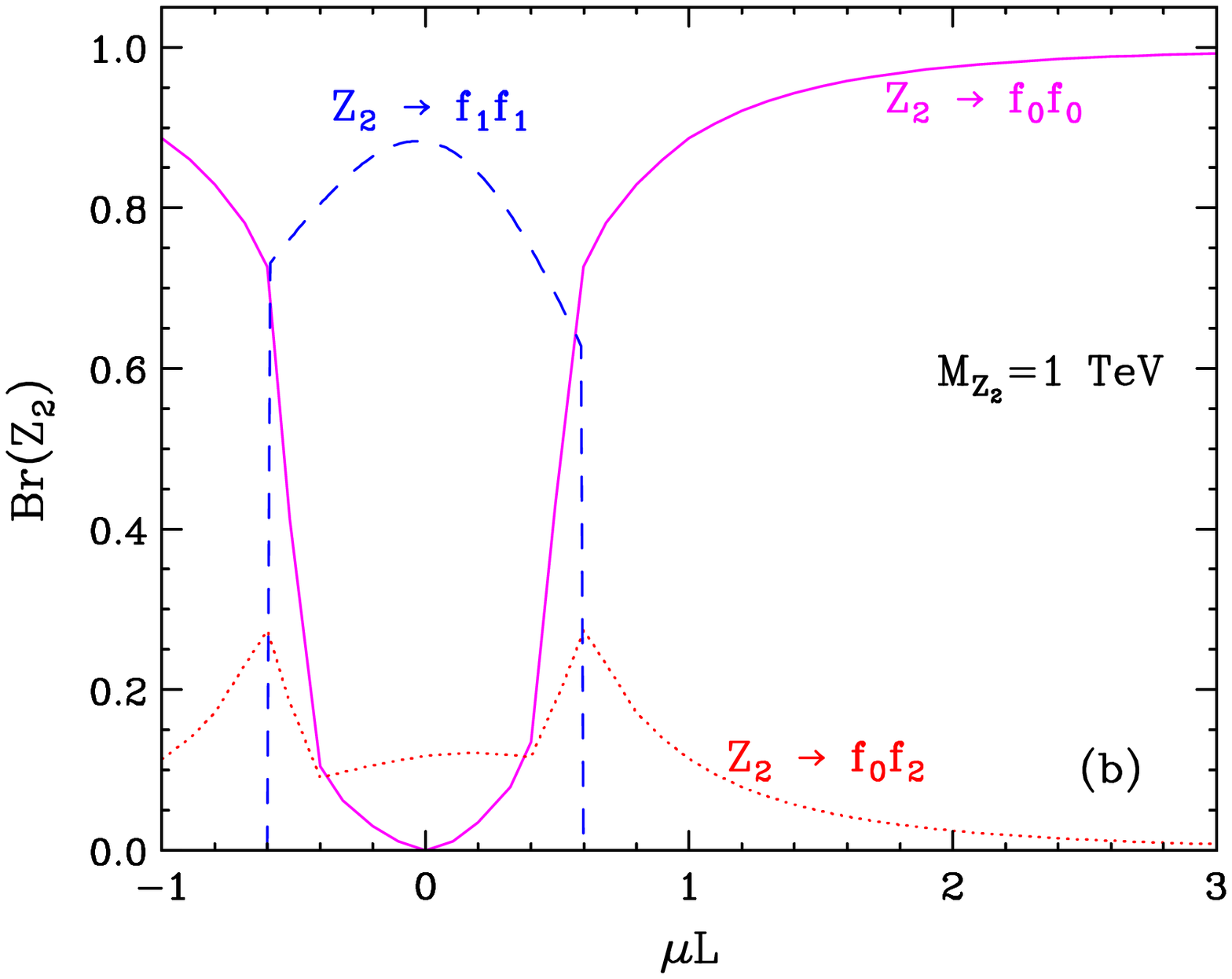,width=7.5cm} }
\caption{\sl (a) The ratio of widths of level-2 KK bosons to their masses and 
(b) branching fractions of 1 TeV $Z_2$, as a function of $\mu L$.}
\label{fig:width_br}}

The 1-loop correction is expected to be the smallest for $\gamma_n$ 
among all other KK states at the same level, and in fact, KK photons receive 
negligible correction from RG running, making the lightest KK photon 
a viable dark matter candidate.
Therefore the decay channels of $\gamma_2$ into $f_1$-$f_1$ or $f_0$-$f_2$ are closed 
and $\gamma_2$ always can appear as a resonance.
As in Eqs. (\ref{eq:width1})-(\ref{eq:width2}), the width dependence on the SM fermion mass 
is negligible even for the top quark, if the resonance is heavy enough.
In this case, the ratio of the total width to its mass becomes 
mass-independent. 
The total width of $Z_2$ ($\gamma_2$) is then $\sim$ 7\% (3.5\%) of its mass 
as $\mu$ increases, as shown in Fig.~\ref{fig:width_br}(a), 
while in MUED, the widths of level-2 KK bosons are 
much less than 1\% \cite{Datta:2005zs}. 
This makes it challenging to probe double resonances which are separated from each other in mass 
by the small 1-loop corrections (7\% or so).

The branching fractions of the $\gamma_2$ are $\mu$-independent for a universal bulk mass, 
which is the case that we consider. They are 25\% in dilepton, 36.7\% in dijet, 
4.2 in $b\bar{b}$, 14\% $t\bar{t}$ and 12.5\% in $\tau \bar\tau$.
The $\gamma_2$ decays invisible through SM neutrinos 7.5\% of the time.
Notice that the branching fraction into dilepton channel is about 20 times larger than 
in case of MUED.

The decay of $Z_2$ is somewhat more complicated than $\gamma_2$ due to the slightly larger 
1-loop correction, which we assume to be about 7\% as in the MUED.
In this case the decay modes to other KK states remain open. 
Without knowing the exact 1-loop mass corrections for all KK particles, 
it is impossible to compute its total width and branching fractions.
For a rough estimate (only for this purpose), we assume that 
KK fermions only gets corrections from the bulk mass 
while $Z_2$ gets heavier by 7\% from RG running.
This is certainly an inconsistent setup.
However, 1-loop corrections to KK fermion masses are known to be merely a few percents 
(1\% for singlet KK fermions and 3\% for doublet KK fermions), and 
for a large $\mu$, bulk mass enhances its mass by quite a large amount and 
this 1-loop contribution becomes negligible. 
This approximation is valid in a broad range of $\mu$.
Given that, one can compute the partial widths of $Z_2$ in three different channels and 
the results are shown in Fig.~\ref{fig:width_br}(b).
The level-2 KK fermion does not get correction from the bulk mass, but the 
$f_0$-$f_2$-$Z_2$ coupling becomes smaller, as shown in Fig.~\ref{fig:couplings}, 
making the relevant branching fraction smaller for large $\mu$.
The same is true for $f_1$-$f_1$-$Z_2$ while $f_0$-$f_0$-$Z_2$ coupling behaves 
in the opposite manner.
Moreover, unlike $f_2$, the level-1 KK modes get heavier as the $\mu L$ increases, and 
at some value of $\mu$, the $Z_2$ decay to $f_0f_2$ gets closed. 
In Fig.~\ref{fig:width_br}(b), this transition value of $\mu L$ is about 0.6 for a 1 TeV $Z_2$.

Having $f_0$-$f_0$-$Z_2$ as a dominant channel, it is straightforward to compute 
the relevant branching fractions. Since $Z_2$ ($W_2^3$) couples to SM pair with the same strength, 
one needs to count relevant degrees of freedom.
The branching fractions are 1/24 in $\tau\bar\tau$, 1/12 in dilepton, 
1/2 in dijet and 1/8 in either $b\bar{b}$ and $t \bar{t}$.
$Z_2$ also can decay invisibly 1/24 of the time.

\subsection{The LHC reach for $\gamma_2$ and $Z_2$ in dilepton channel}
\label{sec:resonances}

We simulate dilepton resonances in the Split-UED at the LHC with 
$\sqrt{s}$=10 TeV, using a private Monte-Carlo generator.
We assume the mass splitting between two bosons ($\gamma_2$ and $Z_2$) is given by 
$M_{Z_2}=1.07  M_{\gamma_2}$ and $M_{\gamma_2}\approx \frac{2}{R}$ as in MUED.
We include both $\gamma_2$ and $Z_2$ in the dilepton signal and 
use the CTEQ6.6 PDF with NLO K-factor.
The leptons from the decay of these KK bosons are highly energetic and 
can easily pass triggers.
For heavy resonances, the energy resolution is better in electron final states 
than in muon final states, and hence we consider electron final state 
with 1\% mass resolution smearing.
$|\eta| < 2.5$ and $M_{\ell\ell} > M_{\gamma_2} - 500$ GeV are imposed as cuts and 
we only count events with dielectron masses greater than 0.8 of $M_{\gamma_2}$.
The dominant background is Drell-Yan, and $t\bar{t}$ and fakes are expected to 
be significantly smaller.
In all cases the background is smaller than the signal by a factor of $\sim$100.

Fig.~\ref{fig:lhc1}(a) shows the required luminosity to observe at least 10 
signal events as a function of $\mu L$ for several values of masses.
The LHC should able to cover the large parameter space 
(up to $M_{V_2} \sim 1.5$ TeV for $\mu L \ge 1$) 
even with the early data at the level of $\sim$100 pb$^{-1}$ or less.
With the integrated luminosity of $\sim$100 fb$^{-1}$, the most of parameter space 
would be probed, setting limit on the bulk mass and the mass of the KK gauge boson.

The expected number of signal events is plotted in the $\mu L$ versus $R^{-1}$ plane 
in Fig.~\ref{fig:lhc1}(b), for ${\cal L}=1$ fb$^{-1}$.
The shaded region in the left side (in yellow) is a projected Tevatron exclusion 
at 95\% C.L. assuming 10 fb$^{-1}$ \cite{CDFdimuon}.
The limit on $R^{-1}$ from $\gamma_2$ gives the best exclusion since it is lighter 
than $Z_2$ and $W_2^\pm$ by 7\%, and 
constrains on $Z_2$ and $W_2^\pm$ are comparable, and hidden below that from $\gamma_2$.
The other shaded area in the left upper corner (in green) is EW constraint from LEP II
considering contact interaction in SUED, as discussed in Section~\ref{sec:constraints}.

\FIGURE[t]{
\centerline{
\epsfig{file=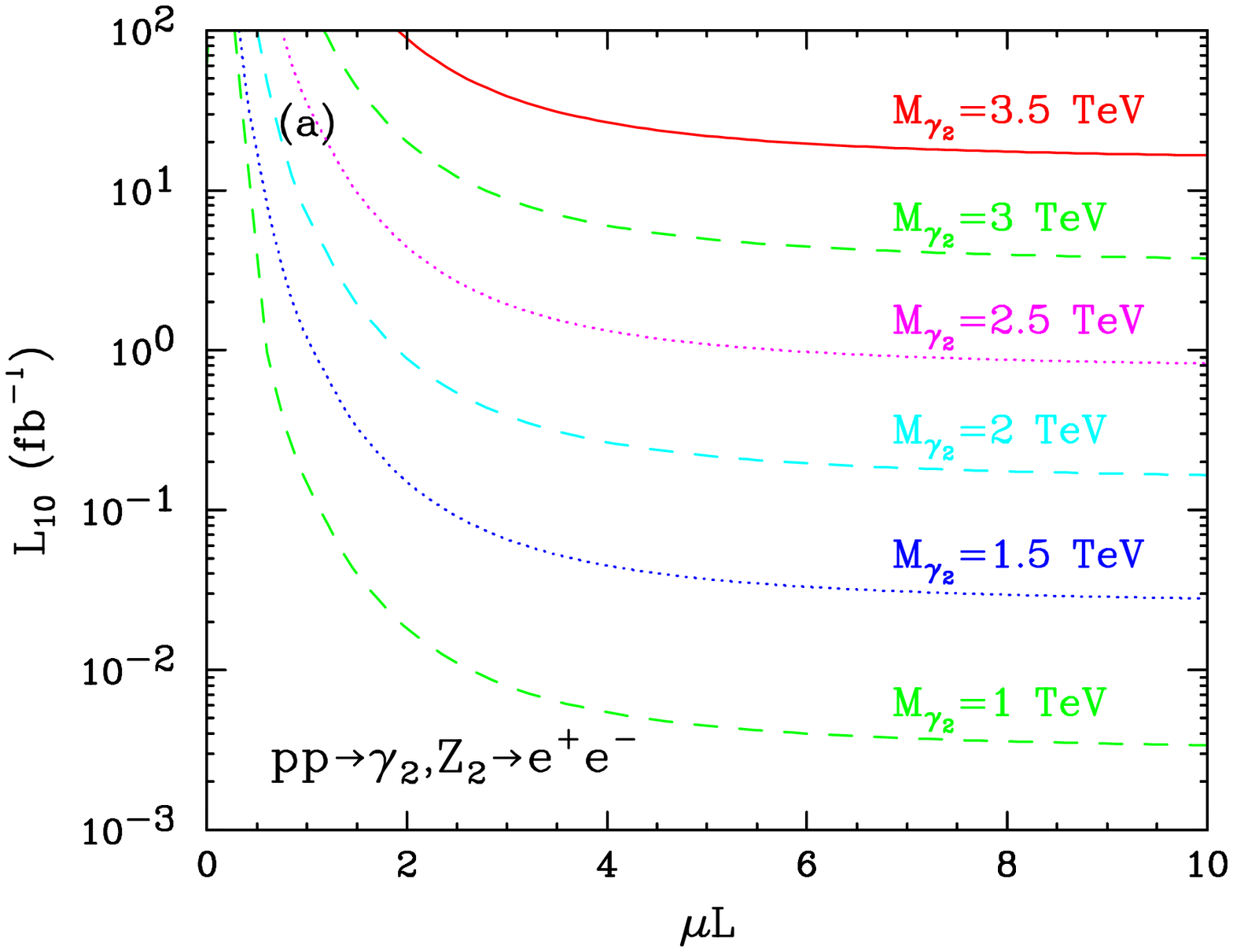, width=7.6cm} \hspace*{-0.2cm}
\epsfig{file=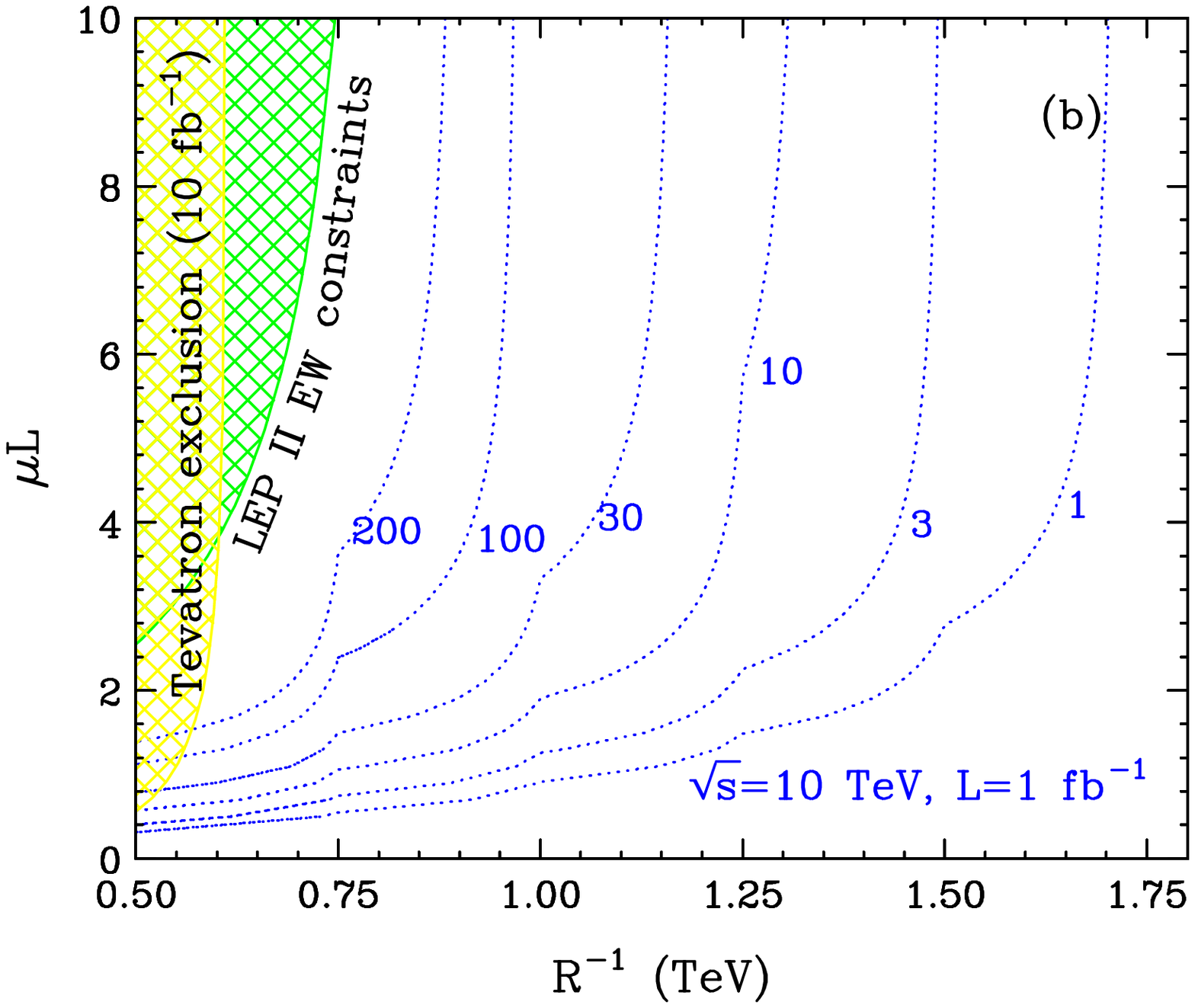,width=7.6cm} }
\caption{\sl The luminosity required to obtain 10 events as a function of $\mu L$ 
for several values of masses in (a) and the number of signal events 
in the $\mu L$ versus $R^{-1}$ plane in (b), 
for $\sqrt{s}$=10 TeV, ${\cal L}=1$ fb$^{-1}$,  $M_{Z_2}=1.07  M_{\gamma_2}$.
In all cases the background is smaller by a factor of $\sim$100. 
We used the CTEQ6.6 with NLO K-factor and 1\% mass resolution smearing.
In obtaining the result we only count events with dilepton masses greater than 0.8 $\times M_{\gamma_2}$.
}
\label{fig:lhc1}}

In Fig.~\ref{fig:invmass} invariant mass distributions are shown for (a) 
$R^{-1}=1$ TeV, $\sqrt{s}= 14$ TeV and ${\cal L}=100$ fb$^{-1}$
and (b) $R^{-1}=0.75$ TeV, $\sqrt{s}= 10$ TeV and ${\cal L}=1$ fb$^{-1}$. 
For both cases, we assume $\mu L \gg 1$.
The yellow histogram is the SM background while 
the red histogram includes both signal and backgrounds.
At the early phase of LHC, one may able to see a bump and get to resolve it into 
double resonance structure as more data gets accumulated.
Notice the negative interference between the SM background and the KK signal, 
which implies the relative sign difference in the couplings. 

%
\FIGURE[t]{
\centerline{
\epsfig{file=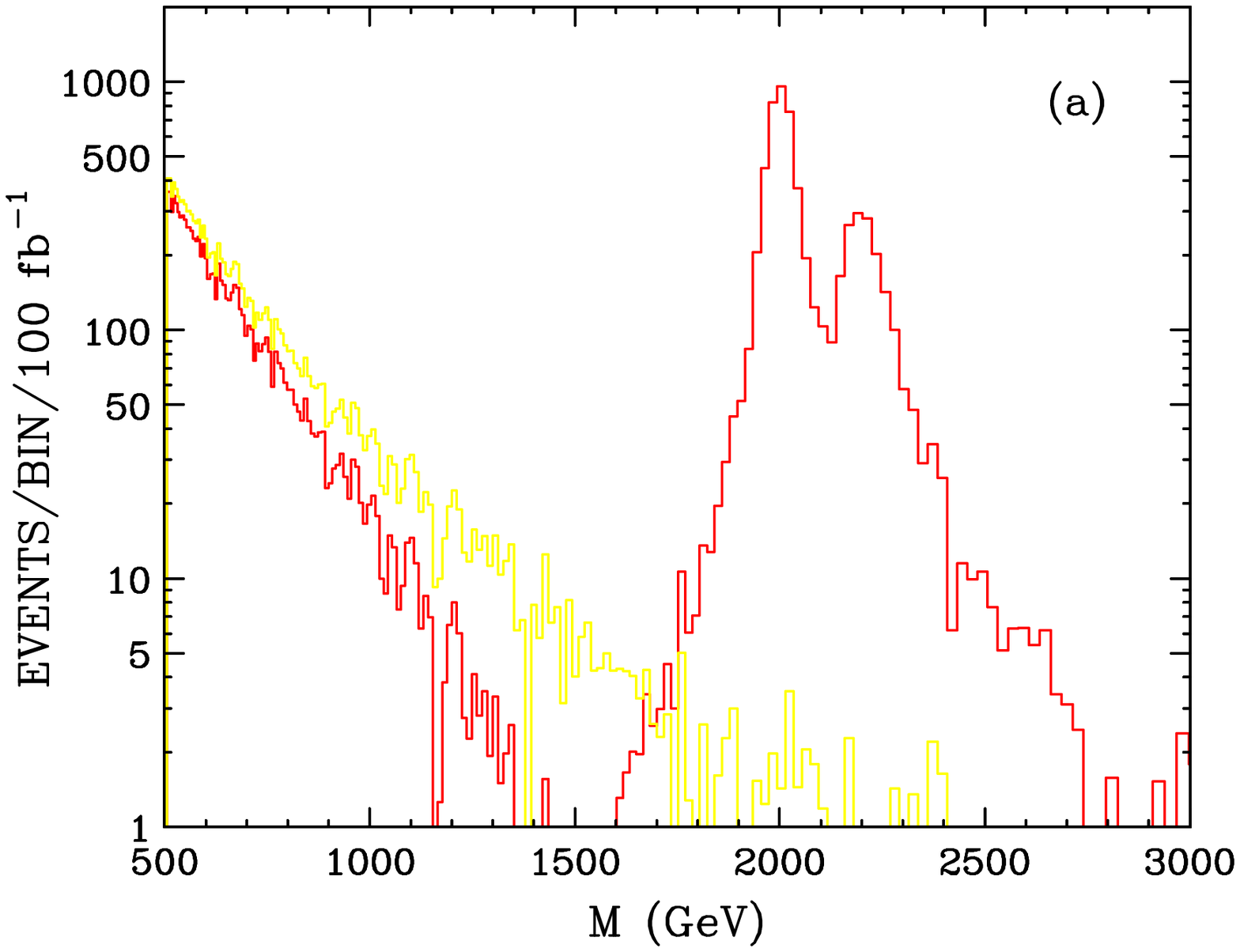, width=7.5cm}\hspace*{0.2cm}
\epsfig{file=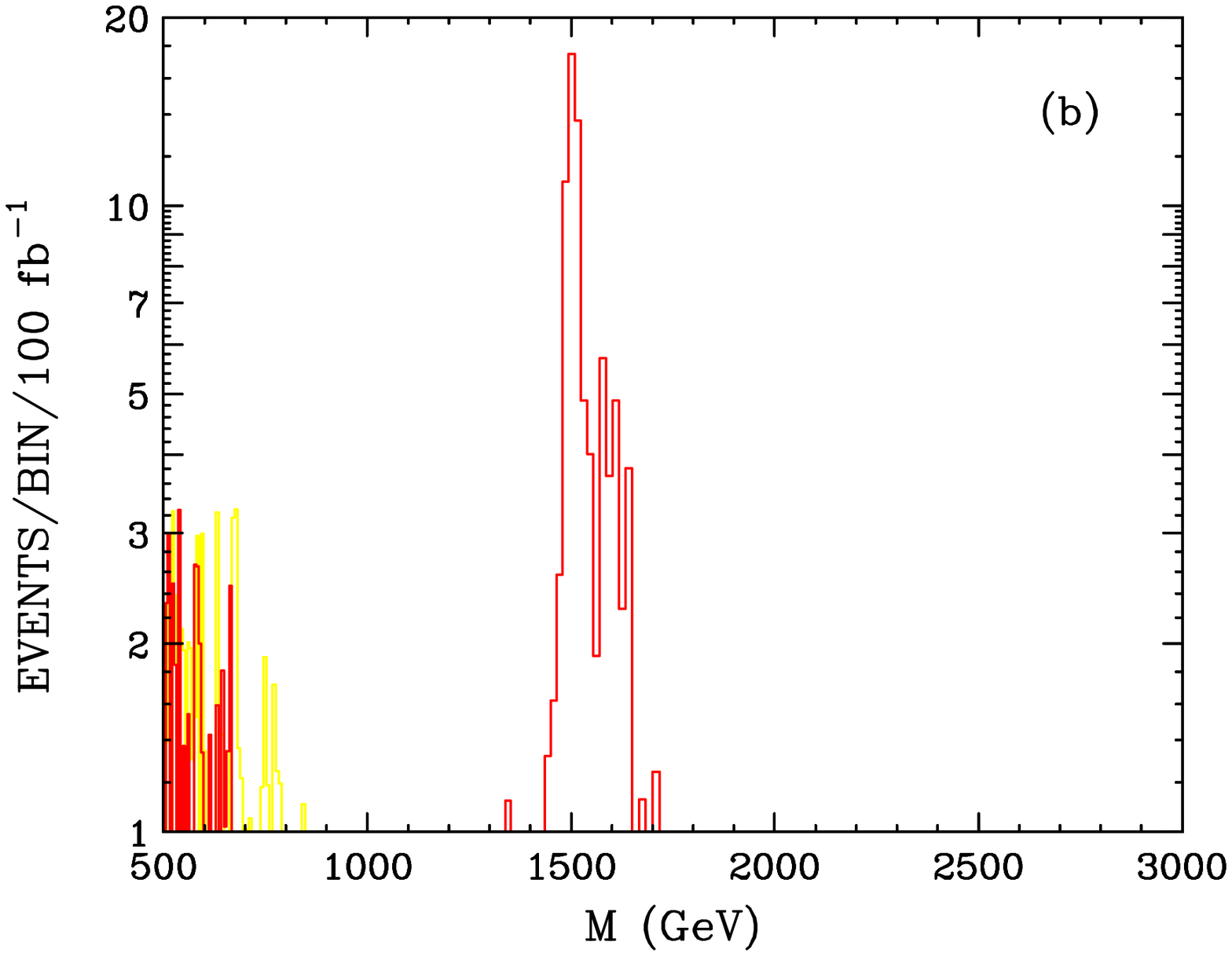,width=7.5cm} }
\caption{\sl Invariant mass distributions at the LHC for (a) 
$R^{-1}=1$ TeV, $\sqrt{s}=14$ TeV and ${\cal L}=100$ fb$^{-1}$
and (b) $R^{-1}=0.75$ TeV,  $\sqrt{s}=10$ TeV and ${\cal L}=1$ fb$^{-1}$.
The yellow histogram is the SM background while 
the red histogram includes both signal and backgrounds.}
\label{fig:invmass}}
%

\section{Conclusions}
\label{sec:conclusion}

The Minimal Universal Extra Dimensions scenario has received great attention.
Recently non-vanishing bulk fermion masses have been introduced 
without spoiling the virtue of KK-parity.
The fermion profiles are no longer simple sine/cosine functions and depend upon 
the specific values of bulk parameters.
The profiles of fermions are split along the extra dimensions 
while the wave functions of the bosons remain the same as in UED.
A simple introduction of a KK-parity conserving bulk fermion mass has significant influences 
on collider aspects as well as astrophysical implications of UED. 
For instance, the DM annihilation fraction into certain SM fermion pairs is 
either enhanced or reduced (compared to the MUED case) so that one can perhaps explain 
the PAMELA positron excess while suppressing the anti-proton flux.

In this paper, we have concentrated on collider phenomenology of 
Split Universal Extra Dimensions. 
We have revisited the KK decomposition in detail and 
analyzed wave function overlaps to compute relevant couplings for collider studies.
We have discussed general collider implication for level-1 KK modes and 
level-2 KK with non-zero bulk mass and have computed LHC reach for the 
EW level-2 KK bosons, $\gamma_2$ and $Z_2$, in the dilepton channel.
The LHC should able to cover the large parameter space 
(up to $M_{V_2} \sim 1.5$ TeV for $\mu L \ge 1$) 
even with early data assuming $\sim$100 pb$^{-1}$ or less.

The existence of double resonances is one essential feature arising from extra dimensional models.
Whether or not one can see double resonances depends 
both on how degenerate the two resonances are and on the mass resolution of the detector.
The very high $P_T$ from the decay makes resolution in dimuon channel worse than in 
dielectron final state. This is because one can reconstruct electron from ECAL but 
muon momentum reconstruction relies on its track, which is barely curved in this case.
Further indication for SUED might be the discovery of 
$W^\prime$-like signature of mass close to $Z_2$.
The MUED predicts a somewhat lower event rate due to 1-loop suppressed 
coupling of level-2 bosons to SM fermion pair, while it exists at tree level in SUED.
Therefore in UED, one has to rely on indirect production of level-2 bosons, whose 
collider study requires complete knowledge of the model: 
{\it the mass spectrum and all the couplings}.
On the other hand, in the large $\mu$ limit of SUED, 
the dependence on mass spectrum is diminished since level-2 KK bosons decay 
only into SM fermion pairs.
This allows us to estimate the signal rate from their direct production, 
so that they can be discovered at the early phase of the LHC. 
The indirect production mechanism only increases production cross sections, 
improving our results.

Once a discovery has been made, one should try to reconstruct events and 
do further measurements such as spin and coupling determination, 
with more accumulated data \cite{Li:2009xh,Petriello:2008zr,Rizzo:2009pu}, 
which might discriminate KK resonances from other $Z^\prime$ models.
The coupling measurement is directly related to the determination of the bulk masses.
A challenging issue might be the existence of two resonances which are rather close to each other.

\bigskip

\acknowledgments
We thank J. Shu and K. Wang for discussion on the forward-backward asymmetry and also 
thank C. Cs\'aki, J. Heinonen and J. Hubisz for helpful discussion.
S. Park is supported by the World Premier International Research Center Initiative 
(WPI initiative) by MEXT and also supported by the Grant-in-Aid for scientific 
research (Young Scientists (B) 21740172) from JSPS, Japan.
K. Kong and T. G. Rizzo are supported in part by the DOE under contract DE-AC02-76SF00515.

\appendix

\section{Appendix: Fermion spectrum in split-UED}
\label{app:spectrum}
\renewcommand{\theequation}{A.\arabic{equation}}
\setcounter{equation}{0}

Let us consider a massive fermion on an orbifold $S^1/Z_2$ with the radius of the circle $R$. 
Two fixed points are at $y=-L$ and $y=+L$ where $L=\pi R/2$. The action is
\begin{equation} 
        \label{eq:BulkAction}
 S = \int d^4 x \int_{-L}^{+L} dy 
\left[i
( \bar{\Psi}\, \Gamma^M {\partial}_M \Psi
 - m_5(y) \bar{\Psi} \Psi  \right] \, ,
\end{equation} 
where  the gamma matrices in 5D  are  $\Gamma^M= (\gamma^\mu, i \gamma_5)$.  

A Dirac  mass is allowed in general. To keep KK-parity, 
the bulk mass must be odd under inversion $m_5(y)=-m_5(-y)$.
We can choose the simplest kink type mass
\begin{eqnarray}
m_5(y) = \mu ~\theta(y) \, ,
\end{eqnarray}
where $\theta(y<0)\equiv -1$ and $\theta(0<y)\equiv +1$.  %
The left (right)-chiral fermion is defined as usual $\gamma_5 \Psi_{L/R}=\mp \Psi_{L/R}$ and 
a generic Dirac fermion is decomposed as $\Psi = \Psi_L +\Psi_R$. Then the action is 
\begin{eqnarray}
S=\int d^4 x \int_{-L}^L dy [&&\bar{\Psi}_L i \gamma^\mu \partial_\mu \Psi_L
+\bar{\Psi}_R i \gamma^\mu \partial_\mu \Psi_R - \bar{\Psi}_L \gamma_5 \partial_5 \Psi_R
-\bar{\Psi}_R\gamma_5\partial_5 \Psi_L \nonumber \\
&&- m_5 (\bar{\Psi}_L \Psi_R +h.c.)] \, .
\end{eqnarray}
Varying the action with respect to $\bar{\Psi}_L$ and $\bar{\Psi}_R$ 
we obtain the standard bulk equations of motion which are given by
\begin{eqnarray}
i \gamma^\mu \partial_\mu \Psi_L - \gamma_5 \partial_5 \Psi_R - m_5 \Psi_R &=&0  \, , \\
i \gamma^\mu \partial_\mu \Psi_R - \gamma_5 \partial_5 \Psi_L - m_5 \Psi_L &=&0 \, ,
\end{eqnarray}
then using $\gamma_5 \Psi_{L/R}=\mp \Psi_{L/R}$ we finally obtain 
\begin{eqnarray}
 \left(\mp \partial_y -m_5\right)\Psi_{R/L} + i \gamma^\mu \partial_\mu \Psi_{L/R} =0 \, .
\end{eqnarray}

\subsection{KK decomposition and wave equations}

Now we would like to discuss how to perform the Kaluza--Klein decomposition of these fields. 
In general, when the fermion belongs to a complex representation of the symmetry group, 
the KK modes can only acquire Dirac masses and the KK decomposition is of the form
\begin{eqnarray} 
        \label{eq:DiracKK}
\Psi_{L/R}=\sum_n \psi_{L/R}^n (x) f_{L/R}^n (y) \, , 
\end{eqnarray} 
where $\psi_{L/R}^n$ are 4D  spinors which satisfy the Dirac equations:
\begin{eqnarray}
i \gamma^\mu \partial_\mu \psi^n_{L/R}=m_n \psi^n_{R/L} \, .
\end{eqnarray}
Plugging this expansion into the bulk equations 
we obtain the following set of coupled first order differential equations 
for the wave functions $f_{L/R}^n$: 
\begin{eqnarray} 
        \label{eq:1st}
\left(\mp \partial_5-m_5\right)f_{R/L}^n + m_n f_{L/R}^n=0 \, .
\end{eqnarray} 

Applying $(\mp \partial_5 + m_5)$ on the first order equations 
we find a decoupled second order equations in the bulk:
\begin{eqnarray} 
0&=&(\mp \partial_5 +m_5)\left[(\mp \partial_5 -m_5)f_{R/L}^n +m_n f_{L/R}^n\right] \nonumber \\
&=&(\partial_5^2 -m_5^2 + m_n^2 \pm m_5')f_{R/L} \\
&=&(\partial_5^2 +\Delta_n^2)f_{R/L}  \nonumber \, ,
 \label{eq:2nd}
\end{eqnarray} 
where $\Delta_n^2 \equiv m_n^2 -m_5^2 \pm m_5'$, where $m_5'$ means $\partial_5 m_5$.

\subsection{Zero mode solution: $m_0=0$}
For $n=0$, we can find a massless solution ($m_0=0$) rather easily. 
In the bulk ($y\neq 0$), Eq. (\ref{eq:1st}) is reduced 
to simple first order equations:
\begin{eqnarray}
\left(\mp\partial_5-m_5\right)f_{R/L}^0=0 \, ,
\end{eqnarray}
having the simple solutions
\begin{eqnarray}
 f_{R/L}^0(y) \sim e^{\mp \int_{-L}^y m_5(y') dy'} \to
 f_{R/L}^0(y) = N_{R/L} e^{\mp \mu |y|} \, ,
 \end{eqnarray}
where the normalization factors are obtained from the condition $\int_{-L}^L |f_{R/L}^0| ^2 =1$:
\begin{eqnarray}
N_{R/L}=\sqrt{\frac{\pm \mu}{1-e^{\mp 2\mu L}}} \, .
\end{eqnarray}

Depending on the sign of $\mu$ the shapes of wave functions are determined. 
If $\mu>0$, e.g., $f_{R}^0$ is localized near the middle point ($y=0$) and 
$f_L^0$ towards the end points ($y=\pm L$). 

\subsection{KK mode solution: Heavy modes ($m_n^2>\mu^2$)}

Depending on the sign of $\Delta_n^2$ the wave functions $f_{R/L}(y)$ 
will be either sines and cosines or hyperbolic sines and hyperbolic cosines. 
Here we first consider the case with $\Delta_n^2=k_n^2>0$. 
In this case the KK modes are heavier than the bulk mass 
since $m_n^2 =\mu^2 + k_n^2>\mu^2$. We call them {\it the heavy modes}.

 The wave equation for the heavy modes looks simple as:
\begin{eqnarray}
(\partial_5^2 + k_n^2)f_{R/L}^n=0 \, ,
\end{eqnarray}
and their generic solutions are
\begin{eqnarray} 
        \label{eq:wv1}
f_{R/L}^n(y) = \alpha_{R/L}^n \cos k_n y + \beta_{R/L}^n \sin k_n y \, .        
\end{eqnarray} 

If $\alpha$'s are zero, $f_{R/L}\propto \sin k_n y$ and 
thus the Dirichlet boundary conditions are reduced to a simple relation:%
\begin{eqnarray}
k_n =\frac{n\pi}{L}=\frac{2 n}{R}, \,\,  n \in {\cal Z} \, .
\label{Eq:even}
\end{eqnarray}
This equation determines the spectrum of even excitation modes. 

If $\alpha$'s are non-zero, on the other hand, 
we get a different set of master equation for Kaluza-Klein spectrum. 
First, $\alpha$'s and $\beta$'s are related by Eq. (\ref{eq:1st}):
\begin{eqnarray} 
&&\mp \alpha_{L/R}^n k_n - m_5 \beta_{L/R}^n+m_n \beta_{R/L}^n=0 \, , \\
&&\pm \beta_{L/R}^n k_n -m_5 \alpha_{L/R}^n + m_n \alpha_{R/L}^n=0 \, ,
\end{eqnarray} 
From the continuity condition at the origin
(${\rm lim}_{\epsilon \to 0}\left( f_{L/R}^n(-\epsilon) -f_{L/R}^n (+\epsilon)\right)=0$) 
we get a useful formula:
\begin{eqnarray}
\mu = \frac{\pm k_n (\beta_{L/R}^{n,>}-\beta_{L/R}^{n,<})}{2 \alpha_{L/R}^n} \, ,
\label{continuity_heavy}
\end{eqnarray}
where we have used $\alpha^>=\alpha^< = \alpha$ from the continuity condition.

Now let us consider boundary conditions. 
We can have two independent choices of Dirichlet boundary conditions 
according to the $Z_2$ orbifold condition: 
$f_L (L) = 0 =f_L (-L)$ (DL) or $f_R(L)=0=f_R(-L)$ (DR).   
\begin{eqnarray}
0&=&f_{L/R}(L)= \alpha_{L/R}^{n,>} \cos k_n L + \beta_{L/R}^{n,>} \sin k_n L \, ,\\
0&=&f_{L/R}(-L)= \alpha_{L/R}^{n,<} \cos k_n L - \beta_{L/R}^{n,<} \sin k_n L \, ,
\end{eqnarray}
or
\begin{eqnarray}
\frac{\beta_{L/R}^{n,>}-\beta_{L/R}^{n,<}}{2 \alpha_{L/R}^n}= - \cot k_n L \, .
\label{BC:D_heavy}
\end{eqnarray} 
Now combining the continuity condition in Eq. (\ref{continuity_heavy}) and 
the Dirichlet boundary condition in Eq. (\ref{BC:D_heavy}) we get the master equation: 
\begin{eqnarray}
\mu = \mp k_n \cot(k_n L)\,\,\, (DL/DR) \, .
\label{Eq:odd}
\end{eqnarray}
This equation determines the KK spectrum for odd excitation modes 
for any given value of $\mu$ (See Fig. \ref{Fig:DLDR}(a) and \ref{Fig:DLDR}(b) 
for DL and DR, respectively).

\FIGURE[t]{
\centerline{
\epsfig{file=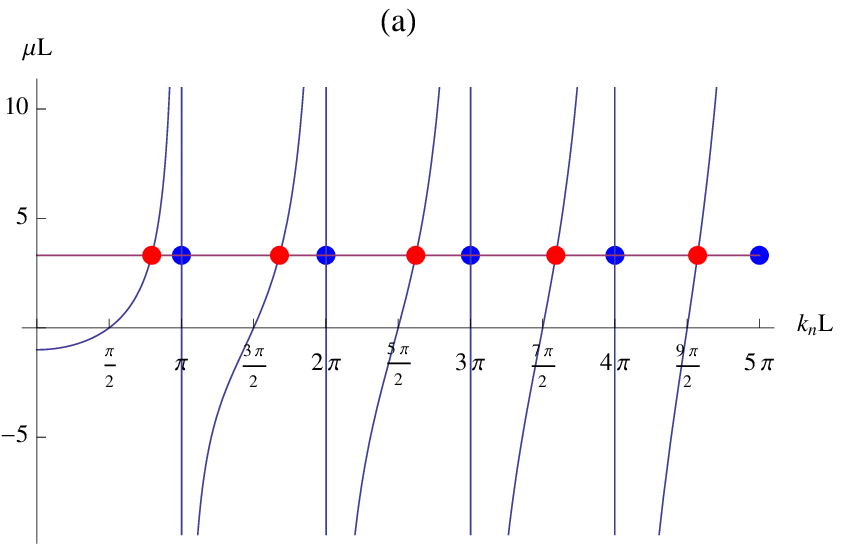,width=.5\textwidth}
\epsfig{file=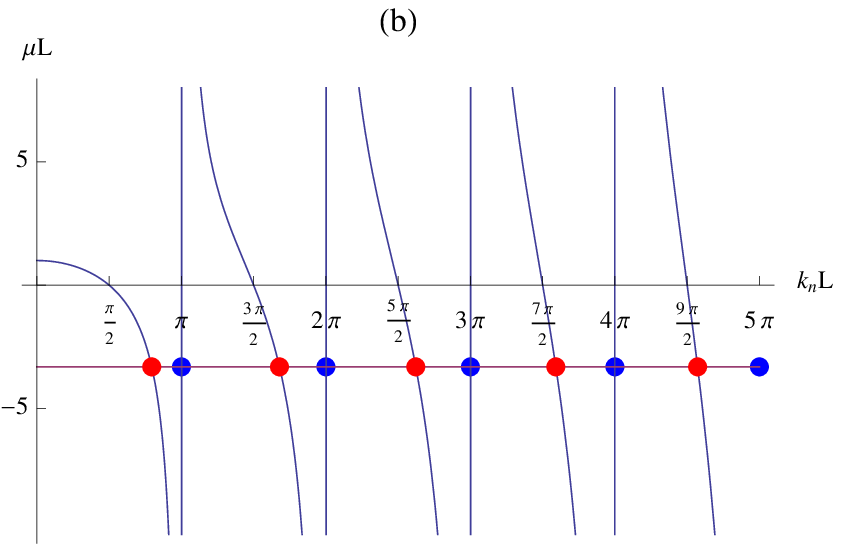,width=.5\textwidth}  }
\caption{(DL) (in (a)) and (DR) (in (b)) Kaluza-Klein spectrum of a fermion with 
a 5D bulk kink-mass with Dirichlet boundary conditions imposed for $\psi_L^{(n)}$'s 
in (a) ($\psi_R^{(n)}$'s in (b)). 
For a given value of $\mu L$, the corresponding $k_n L$ is obtained by 
either $\mu= \mp k_n \cot k_n L$ (odd modes, red dots, Eq.~(\ref{Eq:odd}))
or $k_n L = n \pi$ (even modes, blue dots, Eq.~(\ref{Eq:even}).} %
\label{Fig:DLDR}}

\subsection{KK mode solution: Light mode ($m_n^2<\mu^2$)}
Here we first consider the case with $\Delta_n^2=-\kappa_n^2<0$. 
In this case the KK modes are lighter than the bulk mass 
since $m_n^2 =\mu^2 - \kappa_n^2 < \mu^2$. We call them {\it the light modes}.

 The wave equation for the light modes looks simple:
\begin{eqnarray}
(\partial_5^2 -\kappa_n^2)f_{R/L}^n=0 \, ,
\end{eqnarray}
and their generic solutions are
\begin{eqnarray} 
        \label{eq:wv2}
f_{R/L}^n(y) = \alpha_{R/L}^n \cosh \kappa_n y + \beta_{R/L}^n \sinh \kappa_n y \, .        
\end{eqnarray} 
$\alpha$'s and $\beta$'s are related by Eq. (\ref{eq:1st}):
\begin{eqnarray} 
&&\mp \alpha_{R/L}^n \kappa_n - m_5 \beta_{R/L}^n+m_n \beta_{L/R}^n=0, \\
&&\mp \beta_{R/L}^n \kappa_n -m_5 \alpha_{R/L}^n + m_n \alpha_{L/R}^n=0 \, .
\end{eqnarray} 
From the continuity condition at the origin 
(${\rm lim}_{\epsilon \to 0}\left( f_{L/R}^n(-\epsilon) -f_{L/R}^n (+\epsilon)\right)=0$) 
we get a useful formula:
\begin{eqnarray}
\mu = \frac{\pm\kappa_n (\beta_{L/R}^{n,>}-\beta_{L/R}^{n,<})}{2 \alpha_{L/R}^n} \, .
\label{continuity_light}
\end{eqnarray}
where we have again used $\alpha^>=\alpha^< = \alpha$ from the continuity condition.

Now let us consider boundary conditions. As before we can have two independent choices of 
Dirichlet boundary conditions according to the $Z_2$ orbifold condition: 
$f_L (L) = 0 =f_L (-L)$ (DL) or $f_R(L)=0=f_R(-L)$ (DR).   
\begin{eqnarray}
0&=&f_{L/R}(L)= \alpha_{L/R}^{n,>} \cosh k_n L + \beta_{L/R}^{n,>} \sinh k_n L \, , \\
0&=&f_{L/R}(-L)= \alpha_{L/R}^{n,<} \cosh k_n L - \beta_{L/R}^{n,<} \sinh k_n L \, ,
\end{eqnarray}
or
\begin{eqnarray}
\frac{\beta_{L/R}^{n,>}-\beta_{L/R}^{n,<}}{2 \alpha_{L/R}^n}= - \coth k_n L \, .
\label{BC:D_light}
\end{eqnarray} 
Now combining the continuity condition in Eq. (\ref{continuity_light}) and 
the Dirichlet boundary condition in Eq. (\ref{BC:D_light}) we get the master equation: 
\begin{equation}
\mu = \mp \kappa_n \coth(k_n L)
\end{equation}
for (DL/DR), respectively. 
This equation determines the KK spectrum for light modes. 
In contrast to the heavy modes, light modes can exist only for certain range of 
$\mu$. For (DL/DR), a solution can exist only when $\mu L <-1 (>1)$, respectively 
(See Fig.\ref{fig:DR}). For instance, if $\mu>0$, a light mode solution is not allowed for (DL). 
In this case a zero mode, $f_R^0$, is localized near the center ($y=0$). 
On the other hand, if $\mu<0$, 
a light mode solution is not allowed for (DR) and the zero mode, 
$f_L^0$, is localized toward the center.

\FIGURE[t]{
\centerline{
\epsfig{file=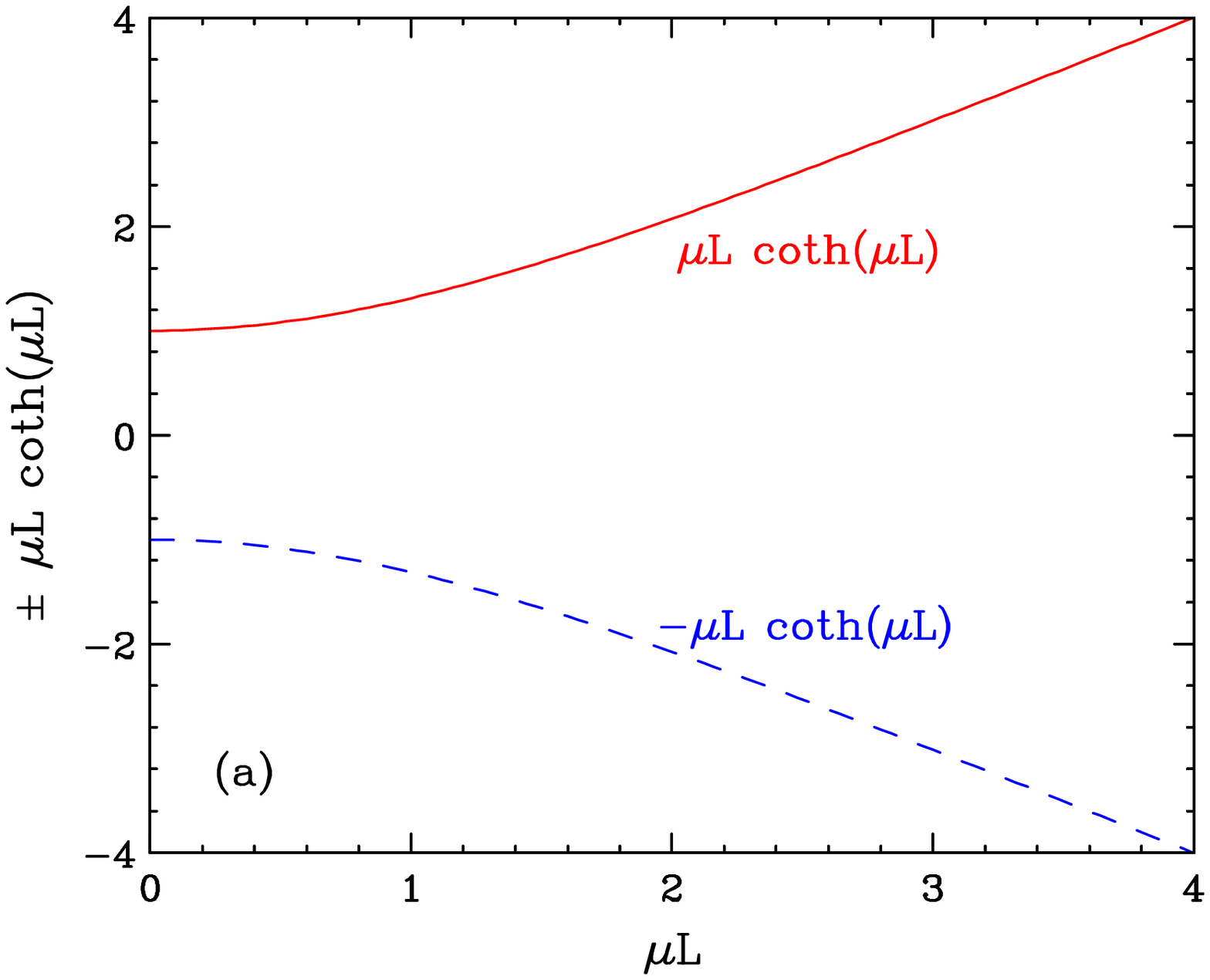,    width=7.5cm}
\hspace{0.1cm}
\epsfig{file=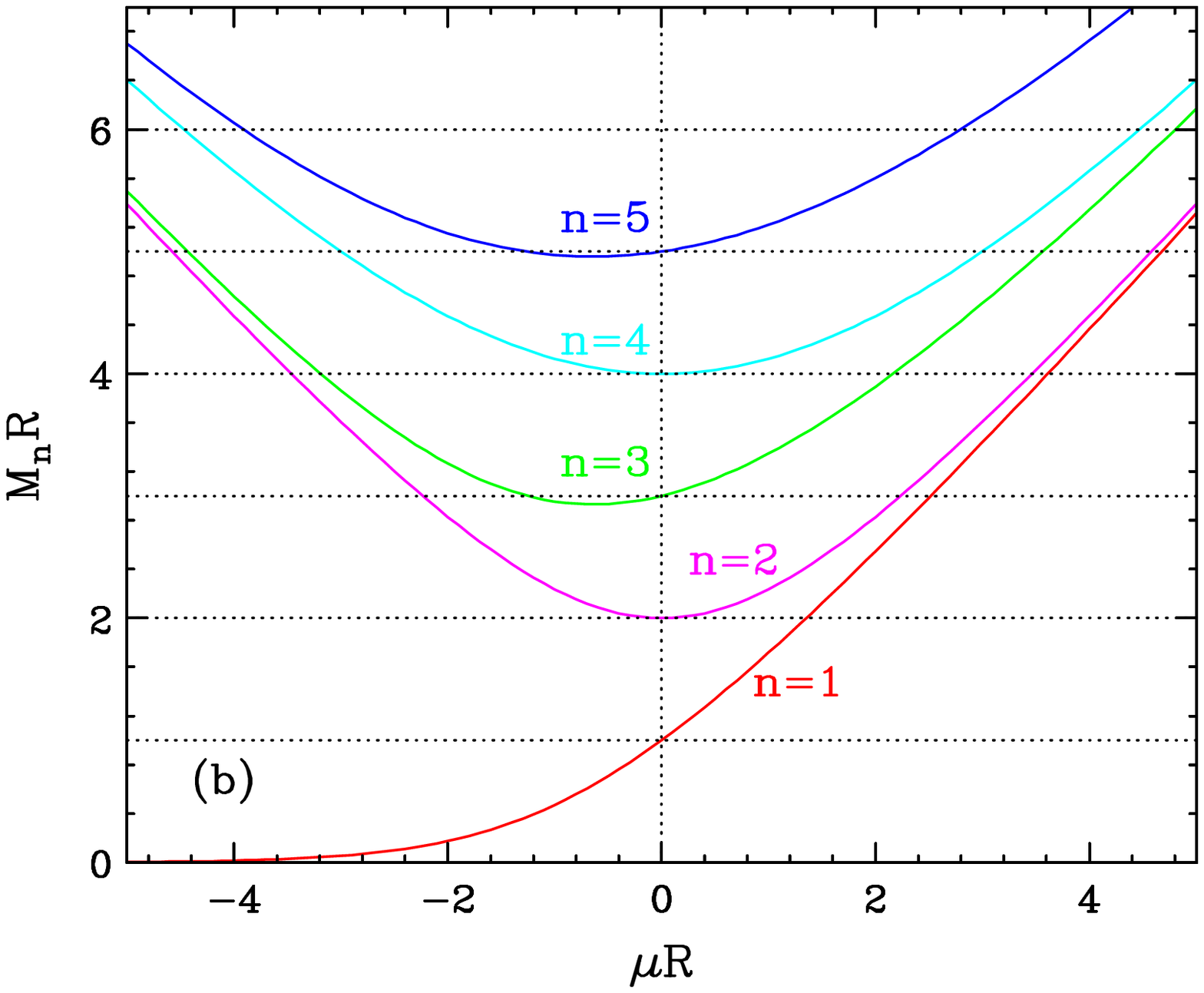, width=7.4cm}}
\caption{(a) Here we plot the function $\pm \kappa L \coth \kappa L$ for (DL/DR), 
respectively. For a given $\mu L$, there exists a unique solution or 
no solution depending on $\mu$. For the (DR) case (upper curve) a solution 
exists when $\mu L >1$. On the other hand, for the (DL) case (lower curve) 
a solution exists when $\mu L<-1$. (b) KK spectrum is obtained in $\mu \in (-5,5)$ 
range for (DL). The lines from the bottom to the top corresponds to 
$n=1,2,3,4,5,\cdots$, respectively. The first KK state becomes very light 
$m_1\sim 2 \mu e^{-|\mu| L}$ as $\mu \to -\infty$. 
The similar spectrum is obtained for (DR) simply by taking inversion $\mu\to-\mu$.
\label{fig:DR}}}

\end{document}